\documentclass[11pt]{article}
\usepackage{graphicx}
\usepackage{latexsym}
\usepackage{amsfonts}
\usepackage{amsmath}
\usepackage{xcolor}
\numberwithin{equation}{section}

\newcommand{\qed}{\hfill $\Box$}

\newtheorem{theorem}{Theorem}[section]
\newtheorem{lemma}[theorem]{Lemma}

\newtheorem{corollary}[theorem]{Corollary}
\newcommand{\be}{\begin{equation}}
\newcommand{\ee}{\end{equation}}
\newcommand{\by}{\begin{eqnarray*}}
\newcommand{\ey}{\end{eqnarray*}}

\newtheorem{example}[theorem]{Example}
\newtheorem{assumption}[theorem]{Assumption}

\newtheorem{remark}[theorem]{Remark}

\begin{document}

\title{ Turnpike Property and Convergence Rate for an Investment Model with General Utility Functions}
\author{Baojun Bian\thanks{Department of Mathematics,
         Tongji University, Shanghai 200092, China.
bianbj@tongji.edu.cn, Research of this author was supported by
NSFC  No. 11371280 and  No. 71090404.}\ \   and
Harry Zheng\thanks{Corresponding Author. Department of Mathematics, Imperial College, London SW7 2BZ, UK. Tel: +44 20 7594 8539. h.zheng@imperial.ac.uk.}}

\date{}
\maketitle

\noindent{\bf Abstract}
In this paper we aim to address two questions faced by a long-term investor with a power-type utility  at  high levels of wealth: one is whether the turnpike property still holds for
a general utility that is not necessarily  differentiable or strictly concave,
the other is whether the error and the convergence rate  of the turnpike property can be estimated. We give positive answers to   both questions. To achieve these results,
we first show that there is a classical  solution to the HJB equation and give a representation of the solution in terms of the dual function of the solution to the dual HJB equation. We demonstrate the usefulness of that representation with some nontrivial examples that would be difficult to solve with the trial and error method.  We then combine the dual method and  the partial differential equation method to give a direct proof to  the turnpike property and to estimate the error and the convergence rate of the optimal policy when the utility function is continuously differentiable and strictly concave.
We finally relax the conditions of the utility function and provide some sufficient conditions that guarantee  the turnpike property and the convergence rate in terms of both primal and dual utility functions.




\newpage

\section{Introduction}
The turnpike property  is a classical problem in financial economics and has been discussed by many researchers for both discrete time and continuous time models,  see Back et al. (1999) and Huang and Zariphopoulou (1999) for exposition and literature.  It is well known that the optimal proportion of wealth invested in the risky asset for a constant relative risk aversion utility is a constant. The turnpike property says the same trading strategy is approximately optimal  at the beginning of the investment period for any utility function behaving asymptotically like a power utility, provided the investment horizon is sufficiently long.  The economic intuition of this phenomenon is that ``When the interest rate is strictly positive, the present value of any contingent claim having payoffs bounded from above can be made arbitrarily small when the investment horizon increases. Thus an investor concentrates his wealth in buying contingent claims that have payoffs unbounded from above at the very beginning of his horizon. As a consequence, it is the asymptotic property of his utility function as wealth goes to infinity that determines his optimal investment strategy at the very beginning of his horizon.'', see Cox and Huang (1992).

 For the Merton problem with a power utility $x^p/p$, where $p<1$ is a constant and $x>0$ is the portfolio wealth, the optimal amount of investment in risky asset at time $t$ is  given by $\theta x/(\sigma(1-p))$, a constant proportion of wealth, where $\theta$ is the Sharpe ratio and $\sigma$  the asset volatility. The optimal amount of investment in risky asset at time $t$ for a general utility $U$ is   given by
\begin{equation}\label{eqn1.0}
A(\tau,x)=- {\theta\over \sigma} \frac{u_x(\tau,x)}{u_{xx}(\tau,x)},
\end{equation}
where $\tau=T-t$ is the time to the investment horizon $T$ and $u$ is the value function that is a solution to a nonlinear partial differential equation (PDE) (see (\ref{eqn18})) and the notation changes above it) with the initial
 condition $u(0,x)=U(x)$, provided that $u$ is continuously differentiable with respect to $\tau$ and $x$. We say the turnpike property holds  if
\begin{equation} \label{eqn1.1}
\lim_{\tau\to\infty} A(\tau,x)={\theta\over \sigma(1-p)} x
\end{equation}
for all $x>0$. The turnpike property (\ref{eqn1.1}) means that the optimal strategy for the utility $U$ is close to the Merton optimal strategy for the power utility  at  any level of the initial wealth $x$ as long as the time to horizon $\tau$ is sufficiently long. Having the turnpike property in portfolio management is highly desirable  as it makes the investment decision process simple and efficient.

One of the standard assumptions in the study of the turnpike property in the literature is that the utility  $U$ is continuously differentiable and strictly concave.
Cox and Huang (1992) use the probabilistic method to show that the  turnpike property holds if the inverse of the marginal utility $(U')^{-1}$  satisfies some conditions. Huang and Zariphopoulou (1999)  establish the turnpike property with the viscosity solution method to the HJB equation when the marginal utility $U'$ behaves like that of a power utility at high levels of wealth and satisfies some other conditions. Jin (1998) discusses an optimal investment and consumption problem  and shows the turnpike property holds in the  sense of convergence on average
($L^1$ and $L^2$ convergence) when $(U')^{-1}$  is ``regularly varying'' at the origin,
see aforementioned papers  for details and the references therein for other models, mainly discrete time models.

Another noticeable missing feature in the literature is that  there are no discussions on the  convergence rate even if the turnpike property is known to hold, that is, if the following inequality holds
\begin{equation} \label{eqn1.2}
\left|A(\tau,x)-{\theta\over \sigma(1-p)} x\right|\leq D(x)e^{-c\tau}
\end{equation}
for some positive constants $c$ and $D(x)$ (see  (\ref{rate})). The significance of (\ref{eqn1.2}) is that it gives  the error estimate of the turnpike property and helps one to determine the  length of the investment period in order to achieve the specified accuracy of replacing the optimal strategy with the Merton optimal strategy.

It is natural and interesting to ask if the turnpike property (\ref{eqn1.1}) still holds for general utilities (strictly increasing, continuous and concave, but not necessarily continuously differentiable and strictly concave) and if the  error estimate  (\ref{eqn1.2}) can be established and  the convergence rate  and the error magnitude  can be computed. Our  main contribution in this paper is to give positive answers to   both questions. The error  and convergence analysis with  closed-form $c$ and $D(x)$ is   the first  in the literature in the study of the turnpike property, to the best of our knowledge.
In the process of proving these results we show the existence of the classical solution to the HJB equation and find the representation of the solution in terms of the dual function, which is of independent interest and may be applied to solve many utility maximization problems  with the stochastic control method.

The discussion of the turnpike property can be decomposed into two problems: one is a finite horizon utility maximization and the other the limiting process for the optimal strategies as the investment horizon tends to infinite.  Stochastic control theory is one of the standard methods for utility maximization. It applies the dynamic programming principle and Ito's lemma to derive a nonlinear PDE, called the HJB equation, for the value function.
If there is a   smooth classical solution to the HJB equation one may use the verification theorem to find the value function and  the optimal feedback control.  For excellent expositions of  stochastic control theory and its applications in utility maximization, see Fleming and Soner (1993) and Pham (2009)
and references therein.

The smoothness of the value function is a highly desirable property as it naturally leads to a feedback optimal control in terms of the value function and its derivatives, which is especially relevant to the turnpike property as the limiting behavior of the optimal strategies is to be studied.   However, one cannot  expect to have a classical solution to the HJB equation unless  some conditions are imposed, for example, the uniform ellipticity of the diffusion coefficient of wealth process, which is not satisfied in general. It is well known that the value function has a  closed-form solution to the HJB equation for the power utility.
When the utility  is strictly concave, continuously differentiable, satisfying some growth conditions,  and the trading constraint set is a closed convex cone, the value function is a classical solution to the HJB equation, see Karatzas and Shreve (1998).
For a general continuous increasing concave utility  $U$ satisfying $U(0)=0$ and $U(\infty)=\infty$,  Bian et al. (2011) show that there exists a classical solution to the HJB equation and that the value function is smooth if an exponential moment condition is satisfied at the optimal control.

To study the turnpike property (\ref{eqn1.1}) and the convergence rate (\ref{eqn1.2}) we first extend the results of  Bian et al. (2011)  to more general utilities. We remove the condition $U(\infty)=\infty$ as we need to address the utility that behaves like  the negative power utility for large wealth.
 We show  that there exists  a classical solution $w$ to the HJB equation and that $w$ has a representation
 $$ w(t,x)=v(t,y(t,x))+xy(t,x)$$
for all  $(t,x)$ in a subset ${\cal S}$ of $[0,T)\times (0,\infty)$,  where $v$ is a smooth solution to the dual HJB equation and $y(t,x)$ is a solution to the equation $v_y(t,y)+x=0$ (see  Theorem \ref{maintheorem}).
We then verify that $w$ is indeed the value function $u$ if a boundedness condition on solution or an exponential moment condition on control is satisfied  (see Theorem  \ref{verification}).

 We illustrate the dual value function technique with several examples. The first one is the Merton problem and shows the value function can be derived without using the trial and error method (see Example \ref{ex0}). The second one is a
terminal wealth maximization problem up to a threshold level, which
  is studied in Xu (2004) in a complete market model with the martingale method. We derive the same result   by applying Theorem \ref{maintheorem} to a specific utility function $U(x)=x\wedge H$ (see Example \ref{ex1}). The third one is a  turnpike problem  with a complicated utility function (see Example \ref{ex4}). The value functions in Examples  \ref{ex1} and  \ref{ex4} are nontrivial  and would be difficult to guess  their forms  without using the suggested solution procedures.

Equipped with the existence of a smooth solution $u$ to the HJB equation and its dual representation with the solution $v$ to the dual HJB equation, we study the turnpike property and the convergence rate with the PDE approach.  The value function $u$ is a solution to  a nonlinear PDE (see (\ref{eqn18})) and is difficult to estimate. We  rewrite $A(\tau,x)$ equivalently as
$$ A(\tau,x)={\theta\over \sigma} yv_{yy}(\tau,y), $$
where $y=u_x(\tau,x)$ and $v$ is a solution to  a linear PDE (see (\ref{eqn20})) and is relatively easy to  analyze as $v$ has a representation in terms of  $V$, a continuous decreasing convex dual function of  $U$ (see Remark \ref{rk3.2}). This transformation is crucial  in estimating the  convergence rate based on that of  $U$.

Since it is more amenable to working with the dual value function  than with the primal value function, we  impose some sufficient conditions on $V$ to ensure the turnpike property and the convergence rate and then show these conditions  are satisfied when  $U$ behaves like a power utility at the large wealth level (see Corollary \ref{corollary3.7}). Theorem \ref{Main-1} is the main result of the paper. It states that if $V$ is continuously differentiable and satisfies
$$ \lim_{y\to 0} {V'(y)\over y^{q-1}} = -1$$
 for some $q<1$
 (see (\ref{V-C}) and Remark \ref{remark3.6}), then  the turnpike property (\ref{eqn1.1}) holds for
 $p=q/(q-1)$ (see (\ref{turnpike})). If, in addition, $V$ satisfies
$$\left|{V'(y)\over y^{q-1}} +1\right|\leq Ky^{\alpha_1}$$
 for some positive constants $K, \alpha_1$ when $y$ is near 0 (see  (\ref{V-C2})), then the convergence rate (\ref{eqn1.2}) holds (see  (\ref{rate})). 

The assumption of the differentiability of $V$  can be removed. The turnpike property (\ref{eqn1.1}) holds if $V(y)$ behaves like $-y^q/q$ for some $q<1$ when $y$ is near 0, or equivalently, $U(x)$ behaves like $x^p/p$ for some $p<1$ when $x$ is very large. The convergence rate (\ref{eqn1.2}) holds if the above limiting behavior is further strengthened by some growth conditions.  These results are proved essentially with the help of Theorem \ref{Main-1} and subdifferential calculus of convex analysis.

The rest of the paper is organized as follows.  In Section 2 we prove the existence of  a
classical solution to the  HJB equation and give the representation of the solution (Theorem \ref{maintheorem}) and  the verification theorem  (Theorem \ref{verification}) with  some examples, including the Merton problem (Example \ref{ex0}) and a wealth maximization problem (Example \ref{ex1}).
In Section 3 we discuss the turnpike property and the convergence rate under the assumption that  the dual utility   is continuously differentiable and satisfies some growth conditions. We illustrate the main result (Theorem \ref{Main-1}) with a nontrivial example (Example \ref{ex4}). In Section 4 we relax the differentiability condition of the dual utility  and prove some sufficient conditions that guarantee  the turnpike property and the convergence rate in terms of both primal and dual utility functions. We also give examples (Remark \ref{remark4.8} and Example \ref{ex5}) to show when the turnpike property may fail to hold. In  Section~5 we  conclude.

\section{Smooth HJB Solutions and Value Functions}
Consider a financial market consisting of one bank account  and $n$ stocks.
The  price process  $S=(S^1,\ldots,S^n)^T$ of $n$ risky assets
is  modeled by
$$
dS_t={\rm diag}(S_t)(\mu(t) dt+ \sigma(t) dW_t),\quad 0\leq t\leq T
$$
with the initial price $S_0=s$,
where  $x^T$ is the transpose of $x$, ${\rm diag}(S_t)$ is an $n\times n$ matrix with diagonal elements $S^i_t$
and all other elements zero,
$\mu$ and $\sigma$ are deterministic continuous
vector-valued and nonsingular matrix-valued functions of time $t$,
representing stock returns and volatilities, respectively,  and
 $W$ is  an $n$-dimensional standard Brownian motion
 on a complete probability space
$(\Omega,\mathcal{F},P)$, endowed with a natural filtration
$\{\mathcal{F}_t\}$ generated by $W$, augmented by all $P$-null sets.
The riskless interest rate is a positive constant denoted by $r$.
The  wealth process $X$  satisfies the stochastic differential equation (SDE)
\begin{equation} \label{wealth}
dX_t = X_t((\pi_t^T b(t) +r)dt + \pi_t^T\sigma(t) dW_t),
\quad X_0=x,
\end{equation}
where $b(t)=\mu(t)-r{\mathbf 1}$ is the stock excess returns, ${\bf 1}$ is a vector with all components 1,  $\pi$ is a progressively measurable
control process satisfying    $\pi_t\in {\cal K}$, a closed convex
    cone in $R^n$, a.s. for $t\in[0,T]$ a.e. $\pi_t$ represents the proportion of wealth $X_t$ invested in risky assets $S_t$. The  process $\pi$  is called admissible if the corresponding wealth process  $X$ is nonnegative for all $t$ a.s. In our notation we write time $t$ in parentheses for deterministic functions (e.g., $b(t),\sigma(t)$) and in subscript for stochastic processes (e.g.,  $S_t,\pi_t$).


The utility maximization problem is defined by
\begin{equation} \label{primal}\sup_{\pi} E[U(X_T)]
\mbox{ subject to (\ref{wealth})},
\end{equation}
where $U$ is a utility function satisfying the following conditions.
\begin{assumption} \label{utility}
$U$ is a continuous increasing  concave function on $[0,\infty)$,
satisfying $U(0)=0$ and
\begin{equation}\label{growth}
 U(x)\leq  C(1+x^p),\ x\geq 0
\end{equation}
for  some constants  $C>0$ and $0<p<1$.
\end{assumption}

\begin{remark}{\rm
Compared with Bian et al. (2011) we have removed the  condition  $U(\infty)=\infty$, which is not satisfied for bounded utilities  such as    $U(x)=-e^{-\alpha x}$ with $\alpha>0$ or
$U(x)=x\wedge H$ with $H>0$.  We denote by $C$ a generic positive constant.
 Since all other conditions are the same as those in Bian et al. (2011) most results in that paper still hold in the current setting. We  state the key results but only prove the parts which are different and refer the reader to  Bian et al. (2011)  for detailed proofs of all other parts. $U(0)=0$ can be replaced by $U(0)>-\infty$.
}\end{remark}

The dual function of $U$ is defined by
\begin{equation} \label{dual_utility}
 V(y)=\sup_{x\geq 0} (U(x)-xy)
\end{equation}
for $y\geq 0$.
The function $V$ is a continuous decreasing and convex function on $[0,\infty)$, satisfying   $V(\infty)=0$ and
\begin{equation} \label{dualgrowth}
V(y)
\leq C(1+y^{q}),\ y>0
\end{equation}
for some constant  $C>0$ and $q={p\over p-1}<0$.

Denote by $u(t,x)$ the  value function of
(\ref{primal}) for $0\leq t\leq T$ and $x\geq0$, defined by
\[
u(t,x)=\sup_{\pi} E[ U(X_T)|X_t=x].
\]
The  HJB equation is given by
\begin{equation} \label{HJB}
{\partial u\over \partial t} + \sup_{\pi\in {\cal K}}
\{\pi^Tb(t)xu_x + {1\over 2}|\sigma(t)^T\pi|^2x^2 u_{xx}\} + rxu_x
 =0
\end{equation}
for $x>0$ and $0<t<T$ with the terminal condition $u(T,x)=U(x)$, where
${\partial u\over \partial t}$ is the partial derivative of $u$ with respect to $t$, $u_x$ and $u_{xx}$  are similarly defined.

The dual process $Y$ satisfies the SDE
\begin{equation} \label{dualSDE}
dY_t=Y_t(-rdt-(\sigma(t)^{-1}\nu_t+\theta(t))^T dW_t), \quad Y_0=y,
\end{equation}
where $\nu$ is progressively measurable  satisfying $\nu_t\in\tilde{{\cal K}}$, the positive polar cone of ${\cal K}$ in $R^n$,  a.s. for  $t\in [0,T]$ a.e. and $\theta(t)=\sigma(t)^{-1}b(t)$. For any admissible control process $\pi$  the process $X_tY_t$ is a super-martingale and therefore the following budget constraint holds:
$$ E[X_tY_t]\leq xy,\;  0\leq t\leq T.$$

The dual minimization  problem is defined by
$$ \inf_{\nu}E[V(Y_T)].
$$

Denote by $v(t,y)$ the dual value function for $0\leq t\leq T$ and $y\geq0$, defined by
$$v(t,y)=\inf_\nu E[ V(Y_T)|Y_t=y].$$
The dual HJB equation is a linear PDE
\begin{equation} \label{dualHJB}
{\partial v\over \partial t}
+\frac{1}{2}|\hat{\theta}(t)|^2 y^2v_{yy} -ryv_y=0,\
y>0,\; 0\leq t<T
\end{equation}
with the terminal condition $v(T,y)=V(y)$,
where $\hat{\theta}(t)=\theta(t)+\sigma(t)^{-1}\hat{\pi}(t)$ and
$\hat{\pi}(t)$ is the unique
minimizer of $f(\tilde{\pi})=|\theta(t)+\sigma(t)^{-1}\tilde{\pi}|^2$ over $\tilde{\pi}\in \tilde{{\cal K}}$.

\begin{assumption}
\label{parabolicity} $\hat{\theta}$
is continuous on $[0,T]$  and there is a positive
constant $\theta_0$ such that $|\hat{\theta}(t)|\geq \theta_0$
for all $t\in [0,T]$.
\end{assumption}

\begin{remark}\label{rk3.2a}
{\rm
Assumption \ref{parabolicity} is automatically satisfied if all components of $b(t)$ are positive, a natural condition as $b(t)$ represents
the stock excess returns, and ${\cal K}$ is either the whole space $R^n$ (no trading constraints) or the nonnegative part of the whole space $R^n_+$ (short selling constraints). The positive polar cone $\tilde {\cal K}$ is then either $\{0\}$ or $R^n_+$ and the optimal solution $\hat\pi(t)=0$ for all $t$. Therefore $\hat\theta(t)=\theta(t)$ is a nonzero continuous vector-valued function on $[0,T]$ and $\theta_0$ is the minimum value of
$|\theta|$ on $[0,T]$.
}\end{remark}

The solution to (\ref{dualHJB}) has the following representation with the Feynman-Kac Theorem:
\begin{equation} \label{kac}
v(t,y)=E[V(Y_T)=E[V(y\tilde Y)],
\end{equation}
where $\tilde Y=\exp(\int_t^T (-r-{1\over 2}|\hat\theta(s)|^2) ds  -\int_t^T \hat\theta(s)^T dW_s)$.
Alternatively, $v$ can be expressed as
\begin{equation}\label{dual_solu}
 v(t,y) = \int_{-\infty}^\infty  K(t, \ln y; T,\xi) V(e^\xi)d\xi,
\end{equation}
where
$$
 K(t,z;s,\xi)=
{1\over \sqrt{4\pi \alpha(t,s)}} \exp\left(-{1\over 4\  \alpha(t,s)} (-r(s-t)-\alpha(t,s)+z-\xi)^2\right)
$$
and $  \alpha(t,s)=\int_t^s \frac{1}{2}|\hat{\theta}(\eta)|^2 d\eta$
for $0\leq t< s\leq T$ and $x,\xi\in R$,

\begin{lemma} \label{LemhatV}
 $v$ is continuous on $[0,T]\times (0,\infty)$ and
 is a classical solution to (\ref{dualHJB}), satisfying
\[
0\leq v(t,y)\leq C (1+y^{q}),\ t\in[0, T], y>0
\]
for some  constant  $C>0$ depending on $T$. Furthermore, for  $t\in[0, T)$,
 $v(t,\cdot)$ is strictly decreasing, strictly convex and has the following limits when $y\to 0$  and  $y\to\infty$:
\begin{eqnarray*}
v(t,0)&=&V(0)  \label{limits0}\\
v(t,\infty)  &=& 0 \label{limits1}\\
v_y(t,0)&=& e^{-r(T-t)}V'(0) \label{limitsd0}\\
v_{y}(t,\infty)&=&0, \label{limitsd1}
\end{eqnarray*}
where $V'$ is the right directional derivative of $V$.
\end{lemma}

\noindent {\it Proof.} $v$ can be  written as
\begin{eqnarray*} v(t,y)&=& \int_{-\infty}^\infty  K(t, \xi; T,0) V(ye^{-\xi})d\xi.
\end{eqnarray*}

Since $V$ is a continuous decreasing convex function and $ K$ is a fundamental solution to (\ref{dualHJB}), we have that $v$ is  continuous on $[0,T]\times [0,\infty)$,
$C^{1,\infty}$ on $[0,T)\times (0,\infty)$, and decreasing and convex in $y$ for  fixed $t\in [0,T]$, therefore, $v_y(t,y)\leq 0$ and $v_{yy}(t,y)\geq 0$. Furthermore, since $v$
is a classical solution to the linear PDE
$$
{\partial v\over \partial t}
+\frac{1}{2}|\hat{\theta}(t)|^2 y^2v_{yy}-ryv_y=0,\
y>0,\; 0\leq t<T,
$$
Applying the strong maximum principle, see Bian et al. (2011, Lemma 3.5), we can show that  $v(t,y)$ is strictly decreasing and strictly convex  in $y$ for fixed $t\in [0,T)$.
We next show the limiting properties of $v$ as $y\to 0$ and $y\to \infty$ for $t\in [0,T]$.

To show $ v(t,0)=V(0) $  we note that
 $V(ye^{-\xi})$ is nonnegative and  increasing as $y\to 0$,
the Monotone Convergence Theorem (MCT) confirms  the desired limit. Here we have used the relation
$\int_{-\infty}^\infty  K(t, \xi; T,0) d\xi=1$.

To show $v(t,\infty)  = 0$ we may say $y>1$ and get $y^{q}<1$ as $q<0$, which gives  $0\leq V(ye^{-\xi})\leq C(1+e^{-q\xi})$,  the Dominated Convergence Theorem (DCT)  implies that $\lim_{y\rightarrow \infty}v(t,y)=0$.

To show  $v_y(t,0)= e^{-r(T-t)}V'(0) $ we write
$$ {v(t,y+h)-v(t,y)\over h}
=\int_{-\infty}^\infty K(t, \xi; T,0) g(\xi,y, h) d\xi,$$
where $ g(\xi,y,h)=(V(ye^{-\xi}+he^{-\xi}) - V(ye^{-\xi}))/h$.
Since $V$ is convex and decreasing  we have $g(\xi,y,h)$ is decreasing as $h\downarrow
0$ and $g(\xi,y,h)\leq 0$ for all $h\geq 0$. The MCT says that
$$
v_y(t,y) =  \int_{-\infty}^\infty K(t, \xi; T,0)  \lim_{h\downarrow 0}  g(\xi,y, h) d\xi=  \int_{-\infty}^\infty K(t, \xi; T,0) e^{-\xi} V'(ye^{-\xi}) d\xi.
$$
Since $V$ is convex we have $V'$ is increasing  and $V'(ye^{-\xi})\leq V'(\infty)=0$ and $V'(ye^{-\xi})$ is decreasing as $y\downarrow 0$. Applying the MCT again we get
$$v_y(t,0)
=  \int_{-\infty}^\infty K(t, \xi; T,0) e^{-\xi}
\lim_{y\downarrow 0} V'(ye^{-\xi}) d\xi
= e^{-r(T-t)}V'(0).$$

Finally, to show  $v_y(t,\infty)=0$ we note that for $y\geq 1$,
$$ 0\leq {V(ye^{-\xi})\over y} \leq {C(1+e^{-q\xi})\over y}\leq C(1+e^{-q\xi}).$$
Applying the DCT we get
$$ v_y(t,\infty)=\lim_{y\to\infty}{v(t,y)\over y}=
 \int_{-\infty}^\infty K(t, \xi; T,0)  \lim_{y\to\infty} {V(ye^{-\xi})\over y} d\xi
=0.
$$
We have proved all the limits.
 \qed

We  can now construct a classical solution to the HJB equation (\ref{HJB}).
\begin{theorem}
\label{maintheorem}
 Assume ${\cal K}$ is a closed convex cone and
Assumptions \ref{utility} and \ref{parabolicity} hold. Then there
exists a  function  $w\in C^0([0,T]\times
[0,\infty))$ which is a
classical solution to the HJB equation (\ref{HJB}) in the region
$S:=\{(t,x): 0\leq t<T, 0<x<-v_y(t,0)\}$ and has the  representation
\begin{equation}
w(t,x)=\left\{\begin{array}{ll}
v(t, y(t,x)) + xy(t,x), & 0\leq x< -v_y(t,0) \\
v(t,0), & x\geq -v_y(t,0),
\end{array}\right. \label{wtx}
\end{equation}
where $y\in
C^{1,\infty}(S)$ satisfies
\begin{equation} v_y(t, y(t,x))+x=0. \label{ytx}
\end{equation}
For $(t,x)\in S$ the function
$w$ is strictly increasing and strictly concave in $x$ for
fixed $t\in [0,T)$ and satisfies $w(T,x)=U(x)$ and
$0\leq w(t,x)\leq C(1+x^p)$ for some  constant $C$.
Furthermore,  the maximum   in the HJB equation (\ref{HJB}) is achieved at
\begin{equation} \label{control}
\pi^*(t,x)=-(\sigma(t)^T)^{-1}\hat\theta(t){w_x(t,x)\over x w_{xx}(t,x)}\in {\cal K}.
\end{equation}
\end{theorem}

\noindent {\it Proof.} For $(t,x)\in [0,T]\times [0,\infty)$ define
\begin{equation} \label{u}
w(t,x)= \inf_{y> 0}\{v(t,y)+xy\}.
\end{equation}

If $x\geq -v_y(t,0)$ then
$$ v(t,y)\geq v(t,0) +v_y(t,0)y  \geq v(t, 0) -xy$$
Therefore
$y\mapsto v(t,y)+xy$ reaches its minimum in (\ref{u}) at  $y=0$ and $w(t,x)=v(t,0)$.

If $0<x<-v_y(t,0)$ then the minimum is reached at a point $y$ satisfying
(\ref{ytx}).
Let $y(t,\cdot)$ be the inverse function of $-v_y(t,\cdot)$,
i.e.,
$$
-v_y(t,y(t,x))= x,\ y(t,-v_y(t,y))=y,
$$
for fixed $t\in[0, T)$. $y(t,x)$ is well defined on
$S$ from Lemma \ref{LemhatV}.
Since $v\in C^{1,\infty}([0, T)\times
(0,\infty))$ and $v_{yy}(t,y)>0$, the inverse function $y\in
C^{1,\infty}(S)$ by  the Implicit
Function Theorem and
$w(t,x)=v(t, y(t,x)) + xy(t,x)$.

Note that if $x=-v_y(t,0)$ then
$-v_y(t, y(t,x))=-v_y(t,0)$ which implies $y(t,x)=0$. So $w(t,\cdot)$ is continuous at $x=-v_y(t,0)$ (if $v_y(t, 0)$ is finite).

For $0<x<-v_y(t, 0)$, since $y(t,x)>0$ and $v_{yy}(t,y(t,x))>0$ for fixed $0\leq
t<T$, the function $w(t,\cdot)$ is strictly increasing and
strictly concave.  A direct computation yields
$$
{\partial w\over \partial t}
-\frac{1}{2}|\hat{\theta}(t)|^2\frac{w_x^2}{w_{xx}}+rxw_x=0.
$$
We conclude by  Bian et al. (2011, Lemma 3.7).
 that $w$ is a classical solution to the
HJB equation (\ref{HJB}) and the maximums of the Hamiltonian are achieved at $\pi^*(t,x)$ and $c^*(t,x)$.  Furthermore, from Lemma \ref{LemhatV} we have
$0\leq w(t,x)\leq  C(1+x^p)$
for some constant $C>0$.
\qed

\begin{remark}
{\rm If $U$ satisfies $U(\infty)=\infty$ then
$V(0)=\infty$ and $V'(0)=-\infty$.  The case $x\geq -v_y(t, 0)$ cannot happen and $w$ is a classical solution to the HJB equation for $(t,x)\in [0,T)\times (0,\infty)$. If $V(0)<\infty$ we may have $V'(0)=-\infty$ (e.g., $U(x)=-e^{-\alpha x}$, $V(0)=0$ and $V'(0)=-\infty$) or $V'(0)>-\infty$ (e.g., $U(x)=x\wedge H$, $V(0)=H$ and
$V'(0)=-H$).
}\end{remark}

Theorem \ref{maintheorem} confirms that there is a classical
solution $w$ to the HJB equation (\ref{HJB}) for $(t,x)\in S$.  The verification theorem next shows  that the value function  $u$ is indeed a smooth classical solution
to the HJB equation (\ref{HJB}) with the optimal feedback control $\pi^*$.

\begin{theorem} \label{verification}
Let $w$ be given as in Theorem \ref{maintheorem} and $u$ be the value function.
If  $x\geq -v_y(t,0)$, then $u(t,x)=w(t,x)=v(t,0)$ and the optimal control is given by $\pi^*_s =0$ for $t\leq s\leq T$. If $0<x<-v_y(t,0)$ then $u(t,x)\leq w(t,x)$ on $[0,T]\times (0,\infty)$. Furthermore, if SDE (\ref{wealth})
admits a nonnegative strong solution $\bar X$
with the feedback control $\bar\pi$ defined in (\ref{control}) and one of the following two conditions is satisfied:
\begin{enumerate}
\item  (boundedness condition on solution) $w(s,\bar X_s)$ is bounded for $t\leq s\leq T$ a.s.;
\item (exponential moment condition on control) $\bar\pi$ satisfies
$ E\left[\exp\left({1\over 2}\int_0^T |\bar\pi_s^T\sigma(s)|^2 ds\right)\right]< \infty,
$
\end{enumerate}
then $u(t,x)=w(t,x)$  and the  optimal control is given by
$$ \pi^*_s= \bar \pi_s 1_{\{ t\leq s\leq \tau^*\}},$$
where $\tau^*$ is a stopping time defined by
$$ \tau^*=\inf\{s\geq t: \bar X_s\geq -v_y(s,0)\}\wedge T$$
and $1_S$ is an indicator that equals 1 if an event $S$ happens and 0 otherwise.
\end{theorem}

\noindent{\it Proof}.\
For $x\geq -v_y(t,0)$ we must have $V(0)$ finite, If we choose
$\pi^*=0$  then wealth at time $T$ is given by
$$ X_T=e^{r(T-t)}x\geq -e^{r(T-t)} v_y(t,0) = -V'(0).
$$
Note that $U(-V'(0))=\inf_{y\geq 0}(V(y)-V'(0)y)=V(0)=U(\infty)$. Therefore
$$
u(t,x)\geq E[ U(X_T) ]\geq U(-V'(0))=v(t,0).
$$
The inequality $u(t,x)\leq v(t,0)$ is obvious as  $U_1(X_T)\leq V_1(0)$ for all  $X_T$.
We have proved that when $x\geq -v_y(t,0)$ the value function
$u(t,x)=w(t,x)=v(t,0)$ and the optimal control $\pi^*\equiv 0$.

For  $0<x< -v_y(t, 0)$ and any feasible control $\pi$ with the corresponding wealth process $X$ and the initial condition $X_t=x$, define a stopping time
$$  \tau=\inf\{s\geq t: X_s\geq -v_y(s,0)\}\wedge T.$$
Then for $t\leq s<\tau$ we have $X_s<-v_y(s,0)$ and
$$
d w(s,X_s) = ({\partial w\over \partial s} + w_x \pi_s bX_s + {1\over 2}w_{xx}\pi_s^2\sigma^2 X_s^2)ds + w_x \pi_s bX_s dW_s.$$
Since $w$ is a nonnegative solution to the HJB equation (\ref{HJB}) we know the drift coefficient above is nonpositive, which implies $w(s,X_s)$ is a supermartingale on $[t,\tau]$. We have
\be \label{eq_verif1} E[w(\tau, X_\tau)|{\cal F}_t]\leq w(t,x).
\ee
If $\tau=T$ then
$$ E[U(X_T)|{\cal F}_t]=E[w(T, X_T)|{\cal F}_t]\leq w(t,x).$$
If $\tau<T$ then $X_\tau=-v_y(\tau, 0)$. We know the optimal control on the interval $[\tau, T]$ is given by $\pi^*\equiv0$ and the value function is given by $u(\tau,X_\tau)= w(\tau,X_\tau)=v(\tau,0)$. Therefore, for any feasible control $\pi$ with the corresponding wealth $X$ on the interval $[\tau, T]$, we must have
\be  \label{eq_verif2} E[U(X_T)|{\cal F}_\tau] \leq w(\tau,X_\tau).
\ee
Combining (\ref{eq_verif1}) and (\ref{eq_verif2}) we have
$$ E[U(X_T)|{\cal F}_t] =E[E[U(X_T)|{\cal F}_\tau]|{\cal F}_t]
\leq E[w(\tau,X_\tau)|{\cal F}_t]\leq w(t,x).
$$
Since $\pi$ is any feasible control, we have shown $u(t,x)\leq w(t,x)$.

Similarly, for  $0<x< -v_y(t, 0)$ and the feedback control $\bar\pi$ with the corresponding wealth process $\bar X$ and the initial condition $\bar X_t=x$, define a stopping time
$$  \tau^*=\inf\{s\geq t: \bar X_s\geq -v_y(s,0)\}\wedge T.$$
Then $w(s,\bar X_s)$ is a local martingale on $[t,\tau^*]$.

If the boundedness condition on solution is satisfied then $w(s,\bar X_s)$ is a martingale on $[t,\tau^*]$, which gives
\be \label{eq_verif3} E[w(\tau^*, \bar X_{\tau^*})|{\cal F}_t]= w(t,x).
\ee
If we  choose the control $ \pi^*_s= \bar \pi_s 1_{\{ t\leq s\leq \tau^*\}}$
with the corresponding wealth process $X^*$ then, using (\ref{eq_verif3}), we have
$$ E[U(X^*_T)|{\cal F}_t]=w(t,x)$$
no matter $\tau^*=T$ or $\tau^*<T$. This gives us the required conclusion
$u(t,x)=w(t,x)$.

If the exponential moment condition on control is satisfied then
we can apply the localization method and  the uniform integrability to show $w(s,\bar X_s)$ is a martingale on $[t,\tau^*]$ and therefore
 $u(t,x)=w(t,x)$, see the detailed proof in Bian et al. (2011, Theorem 4.1).
\qed

We next give some examples to illustrate Theorems \ref{maintheorem} and \ref{verification}. Assume that the wealth process is given by
\begin{equation} dX_t = X_t(rdt + b\pi_t dt + \sigma \pi_t  dW_t) \label{exwealth}
\end{equation}
with $X_0=x$, where $b=\mu-r$, $r,\mu,\sigma$ are positive constants, $W$ a standard Brownian motion, and $\pi$ a progressively measurable control process.
This  is a special case of
(\ref{primal}) in which  ${\cal K}=R$ and its positive polar cone  $\tilde {\cal K}=\{0\}$.

\begin{example} \label{ex0}{\rm
Assume that $U(x)={1\over p}x^p$, where  $0<p<1$ is a constant.
The dual function of $U$ is given by
$V(y)=-{1\over q}y^q$ for $y> 0$, where $q={p\over p-1}$ is a negative constant.
The  solution to the dual HJB equation (\ref{dualHJB}) is given by
$$ v(t,y)= V(y)\exp\left(({1\over 2}q(q-1)\theta^2-r)(T-t)\right),$$
where $\theta=b/\sigma$.
The solution to equation (\ref{ytx}) is given by
$$ y(t,x)=x^{{1\over q-1}}\exp\left((-{1\over 2}q\theta^2+{r\over q-1})(T-t)\right).
$$
A smooth solution to the HJB equation (\ref{HJB}) is given by (\ref{wtx}):
$$ w(t,x)=v(t, y(t,x))+xy(t,x)
= U(x)\exp\left((-{1\over 2}\theta^2{p\over p-1} +r(p-1)) (T-t)\right).$$
The maximum of the Hamiltonian in the HJB equation is achieved at (see (\ref{control}))
$$ \pi^*(t,x)
={\theta\over (1-p)\sigma}.$$
Substituting  $\pi^*(t,x)$ into equation (\ref{wealth}) we get the wealth process $X_t$ satisfying a linear SDE
$$ dX_t= X_t\left((r+{\theta^2\over 1-p} )dt + {\theta\over 1-p}dW_t\right).$$
There is a strong solution to the SDE above and the exponential moment condition  on control
is satisfied. Theorem \ref{verification} confirms that the value function $u=w$.
}\end{example}

\begin{remark}{\rm
Example \ref{ex0} is the famous Merton optimal portfolio selection problem. To solve the HJB equation (\ref{HJB}) a standard method in the literature is to guess the solution having  the form $w(t,x)=U(x)f(t)$, using the scaling property of the power utility $U$, and then solve an ordinary differential equation to get $f(t)$. With the help of    Theorem \ref{maintheorem} we do not need to guess the solution form and can find the solution directly with the dual control method.

}\end{remark}

\begin{example} \label{ex1}
{\rm
Assume that $U(x)=H\wedge x$, where $H$ is a positive constant.
The dual function of $U$ is given by
$V(y)=H(1-y)$ for $0\leq y\leq 1$ and 0 for $y\geq 1$.
The  solution to the dual HJB equation (\ref{dualHJB}) is given by
\begin{eqnarray*} v(t,y)&=& -Hye^{-r(T-t)}\Phi\left(-{1\over \theta \sqrt{T-t}}\ln y
+{r\over\theta}\sqrt{T-t}-{1\over 2}\theta \sqrt{T-t}\right)\\
&& {} +
H\Phi\left(-{1\over \theta \sqrt{T-t}}\ln y
+{r\over\theta}\sqrt{T-t}+{1\over 2}\theta \sqrt{T-t}\right),
\end{eqnarray*}
where
$\theta=(\mu-r)/\sigma$ and
$\Phi$ is the cumulative distribution function of a standard normal variable. Since
$$ v_y(t,y)=-He^{-r(T-t)}\Phi\left(-{1\over \theta \sqrt{T-t}}\ln y
+{r\over\theta}\sqrt{T-t}-{1\over 2}\theta \sqrt{T-t}\right),$$
the solution to equation (\ref{ytx}) is given by
$$ y(t,x)=\exp\left(-\theta\sqrt{T-t} \Phi^{-1}({x\over H}e^{r(T-t)})+(r-{1\over 2}\theta^2)(T-t)\right).$$
The candidate optimal value function is given by (\ref{wtx}):
$$w(t,x)=\left\{\begin{array}{ll}
H\Phi\left(\Phi^{-1}({x\over H}e^{r(T-t)}) + \theta\sqrt{T-t}\right), & 0\leq x< He^{-r(T-t)},  \\
H, & x\geq He^{-r(T-t)}.
\end{array}\right.
$$
In the region of $\{(x,t): 0< x< He^{-r(T-t)}, 0<t<T\}$,  we know
$w$ is a classical solution to the HJB equation (\ref{HJB}) and
the maximum of the Hamiltonian in the HJB equation is achieved at (see (\ref{control}))
\begin{equation} \label{eqn2.26}
\pi^*(t,x)
={He^{-r(T-t)}\over x\sigma\sqrt{T-t} }\phi\left(\Phi^{-1}({x\over H}e^{r(T-t)})\right).
\end{equation}
Substituting the feedback control $\pi^*(t,x)$ into equation (\ref{wealth}) we get the  wealth process $X^*_t$ satisfies a nonlinear SDE
$$ dX_t= rX_tdt +
{He^{-r(T-t)}\over \sqrt{T-t} }\phi\left(\Phi^{-1}({X_t\over H}e^{r(T-t)})\right)
(\theta dt +dW_t).$$
Define $Z_t=f(t,X_t)$, where $f(t, x)=\Phi^{-1}({x\over H}e^{r(T-t)})$. Ito's lemma implies
$$dZ_t = \left({\theta  \over \sqrt{T-t}}
+{1\over 2(T-t)}   Z_t\right)dt
+{1\over \sqrt{T-t}} dW_t.$$
The solution $Z_t$ is given by
$$ Z_t={1\over \sqrt{T-t}} \left(Z_0 \sqrt{T}+
\theta t  + W_t\right)$$
and $Z_0=\Phi^{-1}({x\over H}e^{rT})$.
Therefore the candidate
 optimal wealth process is given by
$$X^*_t=He^{-r(T-t)}\Phi(Z_t).$$
Since $0\leq w(t,x)\leq H$ for all $t$ and $x$, the boundedness condition on solution
is satisfied. Theorem \ref{verification} confirms that the value function $u(t,x)=w(t,x)$ and $X^*$ is the optimal wealth process.
}
\end{example}

\begin{remark}
{\rm
Note that the optimal wealth $X^*_t$ at time $0<t<T$ is a continuous random variable whereas the optimal terminal wealth $X^*_T$  is a Bernoulli random variable taking values 0 and $H$. We have
$$ P(X^*_T=0) =  \Phi\left(-\Phi^{-1}({x\over H}e^{rT})-\theta \sqrt{T}\right)$$
and
$$ w(0,x)=E[U(X^*_T)]=H\Phi\left(\Phi^{-1}({x\over H}e^{rT})+\theta \sqrt{T}\right).$$
It is easy to check that both the value function $w(0,x)$ and the ruin probability
 $P(X^*_T=0)$ are increasing functions of $H$, which shows that the return and the risk are positively correlated. By varying $H$ we can draw a curve on the plane with one axis the expected wealth and the other axis the ruin probability in the same spirit as the celebrated  mean-variance  efficient frontier introduced by Harry Markowitz in 1952.
 }
\end{remark}

\begin{example}\label{example2.18}
{\rm Assume that
\begin{equation} U(x)=\left\{\begin{array}{ll}
x, & 0\leq x< H\\
H(x/H)^p, & x\geq H,
\end{array}\right. \label{example2.18}
\end{equation}
where $H>0$ and $0<p<1$. $U_p$ is a continuous increasing concave function satisfying $U(0)=0, U'(0)=1$ and $U(\infty)=\infty, U'(\infty)=0$, $U$ is not differentiable at $x=H$ and is not strictly concave on the interval $[0,H]$, and $\lim_{p\downarrow 0} U(x)=x\wedge H$.  We may interpret $U$ as the utility for an investor who wants to maximize the absolute portfolio wealth up to a threshold value $H$ and then a scaled power utility when the portfolio wealth is more than $H$.
The dual function is given by
$$ V(y)=H{1-p\over p} p^{1\over 1-p} y^{p\over p-1} 1_{\{0<y\leq p\}}
+ H(1-y) 1_{\{p<y\leq 1\}}.
$$
Some long but straightforward calculation shows that the solution to the dual HJB equation
is given by
$$
 v(t,y)=
H\bigg(  {p_1\over p} p^{1\over p_1} y^{-{p\over p_1}}e^{\alpha(t)^2p\over 2 p_1^2}
\Phi(-c_2+{\alpha(t) p\over p_1})
+\Phi(c_2)-\Phi(c_1)-y\Phi(c_2+\alpha(t))+y\Phi(c_1+\alpha(t)) \bigg),
$$
where $p_1=1-p$, $c_1={1\over \alpha(t)}\ln y - {1\over 2}\alpha(t)$ and
$c_2=c_1 - {1\over \alpha(t)}\ln p$, and its partial derivative  with respect to $y$ is given by
$$
 v_y(t,y)
=H\left( \Phi(c_1+\alpha(t)) - \Phi(c_2+\alpha(t))
-(y/p)^{1\over p-1}e^{\alpha(t)^2 p\over 2(p-1)^2}\Phi(-c_2+{\alpha(t) p\over 1-p})\right).$$
Finally, we can construct a smooth solution $w$ to the HJB equation by
$$ w(t,x)=v(t,y(t,x)) + x y(t,x),$$
where $y(t,x)$ is the unique solution to the equation
$ v_y(t,y)+x=0$.
}\end{example}


\section{Turnpike Property and Convergence Rate}
 In this section we discuss the turnpike property when $T\rightarrow \infty$.
For the Merton problem with a power utility  the optimal strategy is to invest a constant proportion of wealth in the risk asset, called the Merton portfolio. It is in general difficult to find the optimal strategy for a general utility,
 however, if the utility displays the behavior of a power utility when the level of wealth is very high, then  the following Merton strategy can still approximately achieve the optimal value, irrespective of the initial wealth  level, as long as the investment horizon is sufficient long.
This is called the turnpike property and is studied in Huang and Zariphopoulou (1999). Here we not only show a
new and simple proof with the duality method, but also give the estimate of the convergence rate.

The Merton problem tells us the optimal strategy is to invest a constant proportion of wealth in the risk asset for a power utility function. It is in general difficult to find the optimal strategy for a general utility satisfying  (\ref{C3}) and (\ref{C31}), which shows the importance if one can establish the turnpike property as one does not have to find the optimal strategy and, by the following Merton strategy, can still approximately achieve the optimal value as long as the investment horizon is sufficient long.

To simplify the discussion and highlight the essential ideas, we assume in the rest of the paper that the market is made up of one riskless asset with interest rate $r$ and one risky asset with price $S$ satisfying $dS=\mu Sdt+ \sigma S dW$, where $W$ is a standard Brownian motion and $r,\mu,\sigma$ are positive constants satisfying  $\mu>r$. Consider the following utility maximization problem:
$$u(t,x)=\sup E[U(X_T)|X_t=x],$$
where $U$ satisfies the following assumption:
\begin{assumption} \label{utilityturnpike}
$U$ is a continuous, concave and strictly increasing  function on $[0,\infty)$,
satisfying $U(0)=0$ and
\begin{equation}\label{growth1}
 U(x)\leq  C(1+x^{\bar p}),\ x\geq 0
\end{equation}
for  some constants  $C>0$ and $0<\bar p<1$.
\end{assumption}

Let $\bar u(\tau,x)=u(t,x)$  with $\tau=T-t$,  time to horizon. We continue to write $u$ instead of $\bar u$ and $t$ instead of $\tau$ in this section and the next with the understanding that $t$ is a time-to-horizon variable. Therefore $t=0$ is the horizon time and $T\to\infty$ is equivalent to $t\to\infty$.

Theorem \ref{maintheorem}  says that $u$ is a classical solution to the following HJB equation
\begin{equation} \label{eqn18}
-{\partial u\over \partial t}-\frac{1}{2}\theta^2\frac{u_x^2}{u_{xx}}+rxu_x=0,\ (t,x)\in R_+\times R_+
\end{equation}
with  $u(0,x)=U(x)$ for $x\in R_+$ and $\theta=\frac{\mu-r}{\sigma}$, and $u(t,\cdot)$ is strictly increasing and strictly concave for fixed $t>0$.
The optimal amount of investment in risky asset is given by
\begin{equation} \label{primal-portfolio}
A(t,x)=- {\theta\over \sigma} \frac{u_x(t,x)}{u_{xx}(t,x)},\ t>0.
\end{equation}

Let $V$ be the dual function of $U$, i.e.,
 $V(y)=\sup_{x\geq 0}(U(x)-xy)$ for $y\geq 0$.
Then $V$ is a nonnegative, continuous, convex and decreasing function on $[0,\infty)$.
The dual function $v(t,\cdot)$ of $u(t,\cdot)$ satisfies
\begin{equation} \label{eqn20}
{\partial v\over \partial t} -\frac{1}{2}\theta^2 y^2 v_{yy}+ryv_y=0,\ (t,y)\in R_+\times R_+
\end{equation}
with $v(0,y)=V(y)$ for $y\in R_+$ and $v\in C^{1,\infty}$. (\ref{primal-portfolio}) can be equivalently written as
\begin{equation}\label{dual-portfolio}
A(t,x)={\theta\over \sigma} yv_{yy}(t,y),\ t>0,
\end{equation}
where $y$ satisfies $v_y(t,y)+x=0$, or  $y=u_x(t,x)$.
We are interested in estimating the difference of the optimal portfolio $A(t,x)$ and the Merton portfolio ${\theta\over \sigma(1-p)}x$ for the power utility  ${1\over p}x^p$ ($p<1$ and log utility $\ln x$ in case $p=0$) when $t\to \infty$. We therefore need to estimate
\begin{equation} \label{primal-daul-relation}
\left| A(t,x)-{\theta\over \sigma(1-p)}x\right|
= {\theta\over \sigma} \left|yv_{yy}(t,y)+(1-q)v_y(t,y)\right|,
\end{equation}
where $q={p\over p-1}$ and $y=u_x(t,x)$.
It is easy to verify that $w=v_y$ and $w=yv_{yy}$
are the solutions to the equation
\begin{equation}
Lw:={\partial w\over \partial t}-\frac{1}{2}\theta^2 y^2 w_{yy}+(r-\theta^2)yw_y+rw=0,(t,y)\in R_+\times R_+.
\end{equation}

\begin{remark} \label{rk3.2}
{\rm (\ref{primal-daul-relation}) is a key relation we use in the proof of Theorem \ref{Main-1} which leads to the turnpike property and the convergence rate for general utilities  in the next section. The importance of (\ref{primal-daul-relation}) is that it is difficult to directly estimate the left side of (\ref{primal-daul-relation}), which is what we are interested, due to $u$ being a solution to a nonlinear PDE but relatively easy to estimate the right side of (\ref{primal-daul-relation}) due to $v$ being a solution to a linear PDE with an explicit representation in terms of $V$, the  dual  function of the utility  $U$.
}
\end{remark}


We first prove a result that will be used for other results.
\begin{lemma} \label{Asy-1}
\
Let
$w\in C^{1,2}(R_+\times R)$ be a solution to equation
\begin{equation}\label{heat}
{\partial w\over \partial t}-a^2 w_{xx}=0, \
w(0,x)=\phi(x).
\end{equation}
Let $\psi(x)=e^{\alpha x}\phi(x)$ with constant $\alpha>0$. Assume that $\phi\in C^1(R)$ and
\begin{equation}\label{w-c}
 \lim_{x\rightarrow -\infty}\frac{\psi'(x)}{e^{qx}}=-1,\ \
|\psi'(x)|  \leq  \left\{
 \begin{array}{lcr}
Ke^{q x}, &\quad x\leq 0,&\\
K,&\quad x\geq 0,  &
\end{array}
 \right.
\end{equation}
for some constants $K\geq 1$ and $q<1$.
Then we have, for $x\geq 0$,
\begin{equation}\label{PA}
|(e^{\alpha x}w(t,x))_x|\leq K L_0(t),\ \
|(e^{\alpha x}w(t,x))_{xx}|\leq K(|q|+\frac{1}{a\sqrt{\pi t}})L_0(t),
\end{equation}
where $L_0(t)=e^{\alpha^2 a^2 t}+e^{(\alpha -q)^2 a^2 t}$.
Furthermore, we have
\begin{equation}\label{NA}
\lim_{x\rightarrow -\infty} \frac{(e^{\alpha x}w(t,x))_x}{e^{(q-\alpha)^2 a^2 t+qx}}=-1,\ \
\lim_{x\rightarrow -\infty} \frac{(e^{\alpha x}w(t,x))_{xx}}{e^{(q-\alpha)^2 a^2 t+qx}}=-q,
\end{equation}
where the convergence is uniform for $t\in[t_0,t_1]$ with any $0< t_0<t_1$.
\end{lemma}

\noindent{\it Proof.}\ By Poisson's formula, we have
$$w(t,x)=\frac{1}{2a\sqrt{\pi t}}\int_{-\infty}^\infty
e^{-\frac{(\xi-x)^2}{4 a^2 t}}\phi(\xi)d\xi.
$$
An easy calculus shows that
\begin{eqnarray}
e^{\alpha x}w(t,x)&=&\frac{1}{2\sqrt{\pi }}\int_{-\infty}^\infty
e^{-\frac{\eta^2}{4}-\alpha a\sqrt{t}\eta}\psi(x+a\sqrt{t}\eta)d\eta \nonumber\\
(e^{\alpha x}w(t,x))_x&=&\frac{1}{2\sqrt{\pi }}\int_{-\infty}^\infty
e^{-\frac{\eta^2}{4}-\alpha a\sqrt{t}\eta}\psi'(x+a\sqrt{t}\eta)d\eta \label{Po3}\\
(e^{\alpha x}w(t,x))_{xx}&=&\frac{1}{2\sqrt{\pi }}\int_{-\infty}^\infty
e^{-\frac{\eta^2}{4}-\alpha a\sqrt{t}\eta}\left(\frac{\eta}{2a\sqrt{t}}+\alpha\right)\psi'(x+a\sqrt{t}\eta)d\eta. \label{Po5}
\end{eqnarray}

To prove the first inequality in (\ref{PA}), noting from (\ref{w-c}) that, for $x\geq 0$,
\[
|\psi'(x+a\sqrt{t}\eta)|\leq K(1+e^{qa\sqrt{t}\eta}),\; \forall \eta\in R
\]
and $\frac{1}{2 \sqrt{\pi }}\int_{-\infty}^{\infty}
e^{-\frac{\eta^2}{4}-A\eta}d\eta=e^{A^2 }$ for a constant $A$, we have
\[
|(e^{\alpha x}w)_x|\leq\frac{K}{2 \sqrt{\pi}}\int_{-\infty}^\infty
e^{-\frac{\eta^2}{4}-\alpha a\sqrt{t}\eta}(1+e^{qa\sqrt{t}\eta})d\eta=K(e^{\alpha^2 a^2 t}+e^{(q-\alpha)^2 a^2 t}).
\]
The second inequality in (\ref{PA}) is
 proved similarly, using
(\ref{Po5}).

Next we prove (\ref{NA}). Since, using (\ref{Po3}),
\begin{equation}\label{e2}
\frac{(e^{\alpha x}w)_x}{e^{(q-\alpha)^2 a^2 t+qx}}+1= \frac{1}{2 \sqrt{\pi}}\int_{-\infty}^\infty
e^{-\frac{(\eta-2(q-\alpha)a\sqrt{t})^2}{4}}\left(\frac{\psi'(x+a\sqrt{t}\eta)}{e^{q(x+a\sqrt{t}\eta)}}+1\right)d\eta
\end{equation}
and, for $x\leq 0$,
\begin{equation}\label{e3}
\left|\frac{\psi'(x+a\sqrt{t}\eta)}{e^{q(x+a\sqrt{t}\eta)}}\right|\leq K(1+e^{-q a\sqrt{t}\eta }),\
\eta\in R,
\end{equation}
the dominated convergence theorem gives
\[
\lim_{x\rightarrow -\infty}\left|\frac{(e^{\alpha x}w)_x}{e^{(q-\alpha)^2 a^2 t+qx}}+1\right|\leq
\frac{1}{2 \sqrt{\pi}}\int_{-\infty}^\infty
e^{-\frac{(\eta-2(q-\alpha)a\sqrt{t})^2}{4}} \lim_{x\rightarrow -\infty}
\left|\frac{\psi'(x+a\sqrt{t}\eta)}{e^{q(x+a\sqrt{t}\eta)}}+1\right|d\eta=0.
\]
This proves the first limit in (\ref{NA}).
Similarly, noting that, using (\ref{Po5}),
\[ \frac{(e^{\alpha x}w)_{xx}}{e^{(q-\alpha)^2 a^2 t+qx}}+q=\frac{1}{2\sqrt{\pi }}\int_{-\infty}^\infty
e^{-\frac{(\eta-2(q-\alpha)a\sqrt{t})^2}{4}}\left(\frac{\eta}{2a\sqrt{t}}+\alpha\right)\left(\frac{\psi'(x+a\sqrt{t}\eta)}{e^{q(x+a\sqrt{t}\eta)}}+1\right)d\eta,
\]
we can prove the second limit in (\ref{NA}) with the dominated convergence theorem.
\qed

The next result gives the estimate of the rate of convergence.
\begin{lemma} \label{Asy-2}\
Let the conditions of Lemma \ref{Asy-1} be satisfied.
Furthermore, assume that
\begin{equation}\label{w-c2}
\left|\frac{\psi'(x)}{e^{qx}}+1\right|\leq K e^{\alpha_1 x},\ x\leq 0,
\end{equation}
for some constant $\alpha_1>0$.
Then we have, for $x\leq 0$,
\begin{equation}\label{NA1}
\left|\frac{(e^{\alpha x}w)_x}{e^{(q-\alpha)^2 a^2 t+qx}}+1\right|\leq
K L_1(t) (e^{\alpha_1 x}+e^{\frac{x}{a\sqrt{t}}}),
\end{equation}
\begin{equation}\label{NA2}
\frac{|(e^{\alpha x}w)_{xx}-q(e^{\alpha x}w)_{x}|}{e^{(q-\alpha)^2 a^2 t+qx}}\leq K L_2(t) (e^{\alpha_1 x}+e^{\frac{x}{a\sqrt{t}}}),
\end{equation}
where
$L_1(t)=e^{(\alpha -q-\alpha_1)^2 a^2 t}+e^{4+\alpha^2 a^2 t}+2e^{4-2(\alpha-q) a\sqrt{t}}$
and
$L_2(t)=(\alpha_1+|q|+\frac{2}{a\sqrt{t}})L_1(t)$.
\end{lemma}

\noindent{\it Proof.}\ Define
\[
N(x)=\frac{1}{2 \sqrt{\pi }}\int_{x}^{\infty}e^{-\frac{\eta^2}{4}}d\eta,\
M(x)=\frac{1}{2 \sqrt{\pi }}\int_{x}^{\infty}|\eta|e^{-\frac{\eta^2}{4}}d\eta.
\]
A simple calculus shows that
\begin{equation}\label{N-M}
N(x),\ M(x)\leq  e^{4-x},\ \forall x.
\end{equation}
We have, noting (\ref{e2}), (\ref{w-c2}) and (\ref{e3}),
\begin{eqnarray*}
\left|\frac{(e^{\alpha x}w)_{x}}{e^{(q-\alpha)^2 a^2 t+qx}}+1\right|
&\leq&\frac{K}{2 \sqrt{\pi }}\int_{-\infty}^{-\frac{x}{a\sqrt{t}}}
e^{-\frac{(\eta-2(q-\alpha)a\sqrt{t})^2}{4}} e^{\alpha_1(x+a\sqrt{t}\eta)} d\eta\\
&& {} + \frac{K}{2 \sqrt{\pi }}\int_{-\frac{x}{a\sqrt{t}}}^\infty
e^{-\frac{(\eta-2(q-\alpha)a\sqrt{t})^2}{4}}(2+e^{-q a\sqrt{t}\eta})d\eta\\
&\leq& Ke^{(\alpha -q-\alpha_1)^2 a^2 t}e^{\alpha_1 x}\\
&& {}+2KN(-\frac{x}{a\sqrt{t}}+2(\alpha -q)a\sqrt{t})
+Ke^{\alpha^2 a^2 t}N(-\frac{x}{a\sqrt{t}}+2\alpha a\sqrt{t})\\
&\leq& KL_1(t) (e^{\alpha_1 x}+e^{\frac{x}{a\sqrt{t}}}).
\end{eqnarray*}
We have used (\ref{N-M}) in the last inequality.
This proves (\ref{NA1}).

Let
\[
B(\eta)=\frac{\eta}{2a \sqrt{t}}+\alpha-q.
\]
Note that
\[
\frac{1}{2 \sqrt{\pi }}\int_{-\infty}^\infty
B(\eta)e^{-\frac{(\eta-2(q-\alpha)a\sqrt{t})^2}{4}}d\eta=0.
\]
We have from (\ref{Po3}) and (\ref{Po5})
\begin{eqnarray*}
\frac{(e^{\alpha x}w)_{xx}-q(e^{\alpha x}w)_{x}}{e^{(q-\alpha)^2 a^2 t+qx}}
&=&\frac{1}{2 \sqrt{\pi }}\int_{-\infty}^\infty
B(\eta)e^{-\frac{(\eta-2(q-\alpha)a\sqrt{t})^2}{4}}\frac{\psi'(x+a\sqrt{t}\eta)}{e^{q(x+a\sqrt{t}\eta)}}d\eta\\
&=&\frac{1}{2 \sqrt{\pi }}\int_{-\infty}^\infty
B(\eta)e^{-\frac{(\eta-2(q-\alpha)a\sqrt{t})^2}{4}}\left(\frac{\psi'(x+a\sqrt{t}\eta)}{e^{q(x+a\sqrt{t}\eta)}}+1\right)d\eta.
\end{eqnarray*}
Hence, for $x\leq 0$, noting (\ref{w-c2}), (\ref{e3}) and (\ref{N-M}), we have
\begin{eqnarray*}
&&\frac{|(e^{\alpha x}w)_{xx}-q(e^{\alpha x}w)_{x}|}{e^{(q-\alpha)^2 a^2 t+qx}}\\
&\leq&   \frac{K}{2 \sqrt{\pi }}e^{\alpha_1 x} \int_{-\infty}^{-\frac{x}{a\sqrt{t}}} |B(\eta)|
e^{-\frac{(\eta-2(q-\alpha)a\sqrt{t})^2}{4}} e^{\alpha_1a\sqrt{t}\eta} d\eta\\
&&{}+\frac{K}{2 \sqrt{\pi }} \int_{-\frac{x}{a\sqrt{t}}}^\infty |B(\eta)|
e^{-\frac{(\eta-2(q-\alpha)a\sqrt{t})^2}{4}}(2+e^{-q a\sqrt{t}\eta})d\eta\\
&\leq& K(\alpha_1+\frac{1}{a\sqrt{\pi t}})e^{(\alpha -q-\alpha_1)^2 a^2 t}e^{\alpha_1 x}
+\frac{K}{a\sqrt{t}}M(-\frac{x}{a\sqrt{t}}+2(\alpha -q)a\sqrt{t})\\
&&{}+\frac{K}{2a\sqrt{t}}e^{\alpha^2 a^2 t}M(-\frac{x}{a\sqrt{t}}+2\alpha a\sqrt{t})
+K|q|e^{\alpha^2 a^2 t}N(-\frac{x}{a\sqrt{t}}+2\alpha a\sqrt{t})\\
&\leq&K L_2(t)(e^{\alpha_1 x}+e^{\frac{x}{a\sqrt{t}}}).
\end{eqnarray*}
This proves (\ref{NA2}).
\qed

Next we give some estimates for the dual value function  $v$.
\begin{lemma} \label{Asy-v}
Assume that $V\in C^1(R_+)$ and satisfies
\begin{equation}\label{V-C}
\lim_{y\rightarrow 0}\frac{V'(y)}{y^{q-1}}=-1,\ \
|yV'(y)|  \leq  \left\{
 \begin{array}{ll}
Ky^q,  &  y\leq 1, \\
K,& y\geq 1,
\end{array}
 \right.
\end{equation}
where $q<1$. Then we have
\begin{equation}\label{PAA}
|yv_y(t,y)|\leq Ke^{\beta t}L_0(t),\ \ |y^2v_{yy}(t,y)|\leq K(1+|q|+\frac{1}{a\sqrt{\pi t}})e^{\beta t}L_0(t),
\end{equation}
for $y\geq 1$ and
\begin{equation}\label{NAA}
\lim_{y\rightarrow 0}\frac{v_y(t,y)}{e^{\lambda t}y^{q-1}}=-1,\ \
\lim_{y\rightarrow 0}\frac{v_{yy}(t,y)}{e^{\lambda t}y^{q-2}}=1-q,
\end{equation}
where the convergence is uniform for $t\in[t_0,t_1]$ with any $0<t_0<t_1$ and
$\alpha=\frac{1}{2}+\frac{r}{\theta^2}$, $a=\frac{1}{\sqrt{2}}\theta$, $\beta=-a^2\alpha^2$,
$\lambda=\frac{1}{2}\theta^2q(q-1)-rq$.
\end{lemma}

\noindent{\it Proof.}\ Let $w(t,x)=e^{-(\alpha x+\beta t)}v(t,e^x)$.
Then $w$ is a solution to equation (\ref{heat}) with the initial condition
$w(0,x)=\phi(x)=e^{-\alpha x}V(e^x)$.
Since $(\alpha -q)^2 a^2=\lambda-\beta$ and
\begin{equation} \label{yvy}
yv_y=e^{\beta t}(e^{\alpha x}w)_x,\ \ y^2v_{yy}+yv_y=e^{\beta t}(e^{\alpha x}w)_{xx},
\end{equation}
applying Lemma \ref{Asy-1} gives (\ref{PAA}) and (\ref{NAA}).
\qed

\begin{remark} \label{remark3.6}{\rm
Note that the first condition in (\ref{V-C}) can be replaced by $\lim_{y\rightarrow 0}y^{1-q} V'(y)=-k$ for some positive constant $k$. If $k\ne 1$ we may simply divide $V$ by $k$ and work on the new $V$. The turnpike property is unchanged due to the invariance of the optimal  trading  strategy to the scaled  objective function. Note also that the second condition in (\ref{V-C}) can be removed. In fact, for any $y_0>0$ and $y\geq y_0$, the convexity and the nonnegativity of $V$ imply that
$$ V(y_0/2)\geq V(y/2)+V'(y/2)(y_0/2-y/2)\geq V'(y_0/2)y_0/2 -V'(y)y/2,$$
which, together with the decreasing property of $V$, gives
$$0\leq -yV'(y)\leq 2V(y_0/2)- V'(y_0/2)y_0$$
for $y\geq y_0>0$. This, coupled with the first condition in (\ref{V-C}), implies the second condition.
}
\end{remark}

\begin{corollary} \label{corollary3.7}
Assume that $U$ is continuously differentiable and strictly concave, and  satisfies
\begin{equation} \label{cor3.6}
\lim_{x\to\infty} xU'(x)^{1-q}=k, \ \
\end{equation}
where $q<1$ and $k$ is a positive constant.
Then  (\ref{V-C}) holds true.
\end{corollary}

\noindent{\it Proof.} Since $U$ is strictly concave and
$V$ is  defined by  (\ref{dual_utility})
we have $V\in C^1$. Denote by $I=(U')^{-1}$. Then $V(y)=U(I(y))-yI(y)$ and $V'(y)=-I(y)$ for $y>0$. The first condition in (\ref{V-C}) is equivalent to
 $\lim_{x\to\infty} xU'(x)^{1-q}=1$ or $k$ if a scaling is used (see Remark \ref{remark3.6}).
\qed

\begin{remark}\label{remark3.7}{\rm
The condition (\ref{cor3.6}) is equivalent to
\[
\lim_{x\to\infty} {U'(x)\over x^{p-1} }= k^{1-p},
\]
where $p=1+1/(q-1)$, which means the marginal utility of $U$ is 
asymptotically proportional to that of a   power utility  $x^p$.
Furthermore,
if $U$  is twice continuously differentiable, then by applying l'H\^opital's rule, we get
$$ k=\lim_{x\to\infty} {U'(x)^{1-q}\over x^{-1}} =
\lim_{x\to\infty} {(1-q)U'(x)^{-q}U''(x)\over -x^{-2}}
= (1-q)k\lim_{x\to\infty} \left(-{xU''(x)\over U'(x)}\right),
$$
which shows that the relative risk aversion coefficient of $U$, defined by
$R(x):=-xU''(x)/ U'(x)$, converges  to $1-p$ asymptotically as wealth increases.
The inverse is not true in general, see the next example.

\begin{example} \label{ex400}{\rm For a constant $0<p<1$, Define a point $\bar x=\exp(1/(1-p))$ and  a function $U: R_+\to R_+$ by
$$ U(x)=\left\{\begin{array}{ll}
{x\over (1-p)e}, & x\leq \bar x,\\
x^p\ln x, & x>\bar x.
\end{array}\right.
$$ 
Then we have 
$$ U'(x)=\left\{\begin{array}{ll}
{1\over (1-p)e}, & x\leq \bar x,\\
x^{p-1}(p\ln x +1), & x>\bar x
\end{array}\right.
$$ 
and 
$$ U''(x)=\left\{\begin{array}{ll}
0, & x< \bar x,\\
x^{p-2}(p(p-1)\ln x + 2p-1), & x>\bar x.
\end{array}\right.
$$ 
It is easy to check that $U$ is a utility function satisfying Assumption \ref{utilityturnpike}. 
The  relative risk aversion coefficient of $U$ is given by
$$ R(x)=\left\{\begin{array}{ll}
0, & x< \bar x,\\
-{p(p-1)\ln x+ 2p-1\over p\ln x+1}, & x>\bar x.
\end{array}\right.
$$ 
Therefore $\lim_{x\to\infty} R(x)=1-p$. On the other hand, for any $q<1$, we have
$$ xU'(x)^{1-q} = x^{(p-1)(1-q)+1}(p\ln x+1)^{1-q}$$
for $x\geq \bar x$, which converges to 0 if $q<p/(p-1)$ and $\infty$ if $p/(p-1)\leq q<1$. 
Therefore there does not exist a $q<1$ such that (\ref{cor3.6}) holds.

}
\end{example}

}
\end{remark}

\begin{lemma} \label{Asy-v-2}
Assume that the conditions of Lemma \ref{Asy-v} are satisfied. Furthermore, assume that
\begin{equation}\label{V-C2}
\left|\frac{V'(y)}{y^{q-1}}+1\right|\leq K y^{\alpha_1 },\ y\leq 1,
\end{equation}
for some positive constants $K$ and $\alpha_1$. Then we have, for $y\leq 1$,
\begin{equation}\label{NAA3}
\left|\frac{v_y}{e^{\lambda t}y^{q-1}}+1\right|\leq
K L_1(t) \left(y^{\alpha_1}+y^{\frac{1}{a\sqrt{t}}}\right),
\end{equation}
\begin{equation}\label{NAA4}
\frac{|yv_{yy}+(1-q)v_y|}{e^{\lambda t}y^{q-1}}\leq K L_2(t)\left(y^{\alpha_1}+y^{\frac{1}{a\sqrt{t}}}\right).
\end{equation}
\end{lemma}

\noindent{\it Proof.}\
The proof is similar to that of Lemma \ref{Asy-v} by applying Lemma \ref{Asy-2}.
\qed

\begin{remark}\label{remark3.9}{\rm
As in Remark \ref{remark3.6}, the condition (\ref{V-C2}) can be replaced by
\[
\left|\frac{V'(y)}{y^{q-1}}+k\right|\leq K y^{\alpha_1 },\ y\leq 1,
\]
for some positive constants $k$, $K$ and $\alpha_1$.
Furthermore if the limit
\begin{equation} \label{dual_limit}
\lim_{y\to 0} y^{-\alpha_1}\left|\frac{V'(y)}{y^{q-1}}+k\right|=K_0
\end{equation}
exists for some positive constants $\alpha_1$, $k$ and $K_0$, then (\ref{V-C2}) holds.
}
\end{remark}

\begin{corollary}\label{Asy-v-3}
Assume (\ref{V-C}) and (\ref{V-C2}) hold, then  for
$\bar{t}=\frac{1}{a^2 \alpha_1^2}$, we have
\begin{equation} \label{NAA11}
|v_{y}(\bar{t},y)+e^{\lambda \bar{t}}y^{q-1}| \leq \left\{
\begin{array}{lcr}
Ly^{q-1+\alpha_1}, &\quad y\leq 1,&\\
L,&\quad y\geq 1. &
\end{array}
 \right.
\end{equation}
\begin{equation} \label{NAA12}
|yv_{yy}(\bar{t},y)+(1-q)v_{y}(\bar{t},y)| \leq \left\{
\begin{array}{lcr}
Ly^{q-1+\alpha_1}, &\quad y\leq 1,&\\
L,&\quad y\geq 1. &
\end{array}
 \right.
\end{equation}
where
\[
L=K\max\{2e^{\lambda \bar{t}}L_1(\bar{t}),2e^{\lambda \bar{t}}L_2(\bar{t}), e^{\beta \bar{t}}L_0(\bar{t})+e^{\lambda \bar{t}}, (2+2|q|+\frac{1}{a\sqrt{\pi \bar{t}}})e^{\beta \bar{t}}L_0(\bar{t})\}.
\]

\end{corollary}

\noindent{\it Proof.}\ The conclusion follows from  Lemmas \ref{Asy-v} and \ref{Asy-v-2}. \qed

Next theorem gives the turnpike property and the convergence rate of the optimal investment to the Merton portfolio,
 which is the main result of this section.

\begin{theorem} \label{Main-1}
Assume that  $V\in C^1(R_+)$ and (\ref{V-C}) holds.
Then  we have,  for  $x>0$,
\begin{equation}\label{turnpike}
\lim_{t\rightarrow \infty}A(t,x)=\frac{\theta}{\sigma(1-p)}x,
\end{equation}
where $p=\frac{q}{q-1}$.
If, in addition, (\ref{V-C2}) holds for a constant $0<\alpha_1\leq 1-q$,  then
\begin{equation}\label{rate}
\left|A(t,x)-\frac{\theta}{\sigma(1-p)} x\right|\leq D(x) e^{-\frac{r\alpha_1}{1-q} t}
\end{equation}
for $t>\bar{t}$, where
\[
D(x)=(\theta L/\sigma)\{e^{r\bar{t}}+2e^{\frac{r\alpha_1}{1-q}\bar{t}} [2+ 2(L+x)e^{-\lambda \bar{t}} + (2L)^{\frac{1-q}{\alpha_1}}e^{\lambda\frac{1-q}{\alpha_1}) \bar{t}}]^{\frac{\alpha_1+q-1}{q-1}} \},
\]
$\bar{t}$ and $L$ are given in Corollary \ref{Asy-v-3}.

\end{theorem}

\noindent{\it Proof.}\ According to (\ref{PAA}), (\ref{NAA}), we get, for any fixed $t_0>0$, that
\[
\lim_{y\rightarrow 0}\frac{v_{y}(t_0,y)+y^{q-1}e^{\lambda t_0}}{y^{q-1}}=0, \
\lim_{y\rightarrow \infty}(v_{y}(t_0,y)+y^{q-1}e^{\lambda t_0})=0.
\]
For any fixed $\epsilon>0$, there is $\delta=\delta(\epsilon)>0$, such that
\begin{equation} \label{eqn3.3a}
|v_{y}(t_0,y)+y^{q-1}e^{\lambda t_0}|\leq \epsilon y^{q-1}+\delta, \ \ \forall y\in R_+.
\end{equation}
Define
\[
w(t,y)=\pm (v_{y}(t,y)+y^{q-1}e^{\lambda t})+\epsilon (y^{q-1}e^{\lambda (t-t_0)}+1)+\delta e^{-r(t-t_0)}
\]
for $(t,y)\in [t_0,t_1]\times  R_+$ with any $t_1>t_0$. Then $w$ satisfies the following equation and   boundary conditions
\begin{equation} \label{max_prin}
\left\{
\begin{array}{l}
{\partial w\over \partial t}-\frac{1}{2}\theta^2 y^2 w_{yy}+(r-\theta^2)yw_y+rw=r\epsilon,\ (t,y)\in  (t_0,t_1)\times R_+,\\
w(t_0,y)>0,\ y\in R_+,\\
\liminf_{y\rightarrow 0}w(t,y)>0,\ \liminf_{y\rightarrow \infty}w(t,y)>0, \ t\in [t_0,t_1].
\end{array}
\right.
\end{equation}
By the maximum principle, we conclude that $w(t,y)>0$ for all $(t,y)\in  [t_0,t_1]\times R_+$, which implies
\begin{equation} \label{eqn1}
|v_{y}(t,y)+y^{q-1}e^{\lambda t}|\leq \epsilon (y^{q-1}e^{\lambda (t-t_0)}+1)+\delta e^{-r(t-t_0)}
\end{equation}
for all $(t,y)\in  [t_0,t_1]\times R_+$. Noting that for any fixed $x>0$ and $y=u_x(t,x)$, we have $v_{y}(t,y)=-x$ and, from (\ref{eqn1}),
\[
|(u_x(t,x))^{q-1}e^{\lambda t}|\leq x +\epsilon ((u_x(t,x))^{q-1}e^{\lambda (t-t_0)}+1)+\delta e^{-r(t-t_0)}.
\]
Taking $\epsilon=\frac{1}{2}e^{\lambda t_0}$, we conclude that
\begin{equation}\label{Es1}
|(u_x(t,x))^{q-1}e^{\lambda t}|\leq C(x)
\end{equation}
for $t\geq t_0$, where $C(x)=2x + e^{\lambda t_0} + 2\delta(\frac{1}{2}e^{\lambda t_0})$.
Similarly, from (\ref{PAA}) and (\ref{NAA}), we get
\[
\lim_{y\rightarrow 0}\frac{yv_{yy}(t_0,y)+(1-q)v_y(t_0,y)}{y^{q-1}}=0, \ \
\lim_{y\rightarrow \infty}(yv_{yy}(t_0,y)+(1-q)v_y(t_0,y))=0.
\]
For any fixed $\epsilon>0$, there is $\delta=\delta(\epsilon)>0$, such that
\[
|yv_{yy}(t_0,y)+(1-q)v_y(t_0,y)|\leq \epsilon y^{q-1}+\delta,\ y\in R_+.
\]
Define
\[
w(t,y)=\pm (yv_{yy}(t,y)+(1-q)v_y(t,y))+\epsilon (y^{q-1}e^{\lambda (t-t_0)}+1)+\delta e^{-r(t-t_0)}
\]
for $(t,y)\in  [t_0,t_1]\times R_+$. Then $w$ satisfies (\ref{max_prin}).
The maximum principle implies
\begin{equation} \label{eqn3.3b}
|yv_{yy}(t,y)+(1-q)v_y(t,y)|\leq \epsilon (y^{q-1}e^{\lambda (t-t_0)}+1)+\delta e^{-r(t-t_0)}
\end{equation}
for all $(t,y)\in  [t_0,t_1]\times R_+$.
For fixed $x>0$, by (\ref{Es1}), we obtain
\begin{eqnarray}
\left|A(t,x)-\frac{\theta}{\sigma(1-p)}x\right|&=&(\theta/\sigma)|u_x(t,x)v_{yy}(t,u_x(t,x))+(1-q)v_y(t,u_x(t,x))|\nonumber\\
&\leq& (\theta/\sigma)(\epsilon ((u_x(t,x))^{q-1}e^{\lambda (t-t_0)}+1)+\delta(\epsilon) e^{-r(t-t_0)})\nonumber\\
&\leq& (\theta/\sigma)(\epsilon (C(x)e^{-\lambda t_0}+1)+\delta(\epsilon) e^{-r(t-t_0)})
\label{turnpike0}
\end{eqnarray}
for $t\geq t_0$. Let $t\to\infty$ and then $\epsilon\to0$, we derive the turnpike property (\ref{turnpike}).

Next assume that (\ref{V-C2}) holds for a constant $0<\alpha_1 \leq 1-q$.
For any fixed $\epsilon>0$, define
$y_{\epsilon}=(\frac{\epsilon}{L})^{\frac{1}{\alpha_1}}$. Let $t_0=\bar{t}$, then (\ref{NAA11}) implies that
\[
|v_{y}(\bar{t},y)+ e^{\lambda \bar{t}}y^{q-1}|
\leq \max \{\epsilon y^{q-1}, Ly_{\epsilon}^{q-1+\alpha_1}, L \}, \  \forall y\in R_+.
\]
Hence (\ref{eqn3.3a}) is satisfied with
\begin{equation} \label{delta}
\delta(\epsilon)=L+L^{\frac{1-q}{\alpha_1}}{\epsilon}^{\frac{q-1+\alpha_1}{\alpha_1}}.
\end{equation}
 Therefore, we obtain the explicit form for the constant $C(x)$
\[
C(x)=2x + e^{\lambda \bar{t}} + 2\delta(\frac{1}{2}e^{\lambda \bar{t}})
=2(L+x) + e^{\lambda \bar{t}} + (2L)^{\frac{1-q}{\alpha_1}}e^{\lambda(1+\frac{1-q}{\alpha_1}) \bar{t}}.
\]

Similarly, (\ref{eqn3.3b}) is
satisfied with $\delta$ given by (\ref{delta}). Let
\[
\epsilon=L \left((C(x)e^{-\lambda \bar{t}}+1)e^{r(t-\bar{t})}\right)^{\frac{\alpha_1}{q-1}}.
\]
We obtain from (\ref{turnpike0}) and $0<\alpha_1<1-q$ that
\begin{eqnarray*}
\left|A(t,x)-\frac{\theta}{\sigma(1-p)}x\right|&\leq&
(\theta/\sigma)\left(2L (C(x)e^{-\lambda \bar{t}}+1)^{\frac{\alpha_1+q-1}{q-1}}e^{\frac{r\alpha_1}{q-1}(t-\bar{t})}+ Le^{-r(t-\bar{t})}\right)\\
&\leq& (\theta/\sigma)(2L (C(x)e^{-\lambda \bar{t}}+1)^{\frac{\alpha_1+q-1}{q-1}}e^{\frac{r\alpha_1}{1-q}\bar{t}}+Le^{r\bar{t}})
 e^{-\frac{r\alpha_1}{1-q}t}.
\end{eqnarray*}
Substituting $C(x)$ into the above inequality, we get $D(x)$ in (\ref{rate}).
\qed

\begin{remark} {\rm It is clear in the proof of Theorem \ref{Main-1} that the positive  interest rate $r$ plays the crucial role in establishing the turnpike property (\ref{turnpike})
and the convergence rate (\ref{rate}).  This point is also highlighted in Back et al. (1999)  that ``it is the growth of the economy as reflected in interest rates
or discount bond prices, not independence, that is critical for the results''.
}
\end{remark}

Theorem \ref{Main-1} states that the  convergence rate is $r\alpha_1/(1-q)$ and the maximum error  is $D(x)$ which can be computed explicitly. For a specific utility function we may derive sharper error estimates than those given in Theorem \ref{Main-1}. The next example  illustrates that point and  shows again the usefulness of the dual method in finding a solution to the HJB equation, which would be difficult with the trial and error method.



\begin{example} \label{ex4}{\rm
 Define
\begin{equation}
 U(x)={1\over 3}H(x)^{-3}+ H(x)^{-1} + xH(x)
\label{example3.12}
\end{equation}
for $x>0$, where
$$ H(x)=\left({2\over -1+\sqrt{1+4x}}\right)^{1/2}.$$
A simple calculus gives
$U'(x)=H(x)$ and
$U''(x)=-\sqrt{2}(-1+\sqrt{1+4x})^{-3/2}(1+4x)^{-1/2}$,
which shows
$U$ is strictly increasing and strictly concave. Furthermore, $H(0)=\lim_{x\to0}H(x)=\infty$, $H(\infty)=0$ and $\lim_{x\to\infty} xH(x)=\infty$, which gives $\lim_{x\to 0}U(x)=0$ (we may define $U(0)=0$), $U(\infty)=\infty$, $U'(0)=\infty$ and $U'(\infty)=0$. Therefore $U$ is a utility function.  Furthermore, the relative risk aversion coefficient of $U$ is given by
$$ R(x)=-{xU''(x)\over U'(x)}= {1\over 4}\left(1+{1\over \sqrt{1+4x}}\right),$$
which shows that $U$ is not a HARA utility. Since $R$ is a decreasing function and has a limit $1/4$ as $x\to\infty$, $U$ represents an investor who will increase the percentage of wealth invested in the risky asset as wealth increases,  which is realistic economic behavior. To get other information, including the turnpike property, we need to do further analysis.

 The dual function of $U$ is defined by $V(y)=\sup_{x\geq 0}(U(x)-xy)$. Since the optimal point $x$ satisfies the equation $U'(x)-y=H(x)-y=0$, we have  $x=y^{-2}+y^{-4}$ and
\[
V(y)={1\over 3}y^{-3}+y^{-1}.
\]
 It is easy to check that  (\ref{V-C}) holds with $q=-3$ and  $K=2$ and  (\ref{dual_limit}) holds with $\alpha_1=2$ and $K_0=1$. Theorem \ref{Main-1} states that the turnpike property (\ref{turnpike}) and the convergence rate (\ref{rate}) hold.

Since the dual value function $v$ is the solution to the linear PDE (\ref{eqn20}), we can transform it into a simple heat equation and then find the closed-form solution. Specifically, if
we let $\alpha=\frac{1}{2}+\frac{r}{\theta^2}$, $a=\frac{1}{\sqrt{2}}\theta$, $\beta=-a^2\alpha^2$, and  $w(t,z)=e^{-(\alpha z+\beta t)}v(t,e^z)$,
then $w$ is a solution to the equation $w_t-a^2w_{zz}=0$
 and has the initial condition
$w(0,z)=e^{-\alpha z}V(e^z)$ with $V(y)=-y^q/q-y^{\bar q}/\bar q$ and $q=-3, \bar q=-1$. We can use Poisson's formula to find the closed-form solution  $w(t,z)$ and then get $v(t,y)$. The solution
to  the linear PDE (\ref{eqn20}) is given by
$$ v(t,y)=-{1\over q}y^qe^{-rqt+{1\over 2}\theta^2 q(q-1)t}
-{1\over \bar q}y^{\bar q}e^{-r\bar qt+{1\over 2}\theta^2 \bar q(\bar q-1)t}.
$$
Substituting $v_y$ and $v_{yy}$  into
(\ref{primal-daul-relation}) we get
\begin{equation} \label{turnpike-example}
\left|A(t,x)-\frac{\theta}{\sigma(1-p)} x\right|=
{\theta\over\sigma} |q-\bar{q}|e^{-r\bar qt+{1\over 2}\theta^2 \bar q(\bar q-1)t}y^{\bar{q}-1}.
\end{equation}
Furthermore, since $y$ is the solution to  $v_y(t,y)+x=0$, we need to solve the equation
\[
-e^{-rqt+{1\over 2}\theta^2 q(q-1)t}y^{q-1}-e^{-r\bar qt+{1\over 2}\theta^2 \bar q(\bar q-1)t}y^{\bar{q}-1}+x=0.
\]
By the choice of   $q=-3, \bar{q}=-1$, we get
\begin{equation} \label{eqn3.39}
y^2=\frac{1}{2x}\left(e^{(r+\theta^2)t}+\sqrt{e^{2(r+\theta^2)t}
+4xe^{3(r+2\theta^2)t}}\right).
\end{equation}
Finally, substituting $y$ into (\ref{turnpike-example}) leads to
\begin{equation} \label{hhh}
\left|A(t,x)-\frac{\theta}{\sigma(1-p)} x\right|={\theta\over\sigma}\frac{4x}{1+\sqrt{1+4xe^{(r+4\theta^2)t}}}\leq {\theta\over\sigma}2\sqrt{x}e^{-({r\over 2} + 2\theta^2)t}.
\end{equation}

We have found $D(x)=(2\theta/\sigma)\sqrt{x}$. Note also that
the convergence rate $c$ in Theorem \ref{Main-1} is equal to
$c=r/2$ and the convergence rate in (\ref{hhh}) is $r/2+2\theta^2$ which is greater than $c$. This shows that by direct computation one may get a sharper convergence rate  than the one given in Theorem \ref{Main-1}.

Furthermore, we  can derive a classical solution to the primal HJB equation by computing $u(t,x)=\inf_{y>0}(v(t,y)+xy)$, which leads to the equation $v_y(t,y)+x=0$ and
\begin{eqnarray*}
 u(t,x)&=& v(t,y)+xy\\
&=& {y\over 3} (y^{-4}e^{3(r+2\theta^2)t}+3y^{-2}e^{(r+\theta^2)t}+3x)\\
&=&{y\over 3} (-y^{-2}e^{(r+\theta^2)t}+x+3y^{-2}e^{(r+\theta^2)t}+3x)\\
&=& {2\over 3}(y^{-1}e^{(r+\theta^2)t} +2xy),
\end{eqnarray*}
where $y>0$ is given by (\ref{eqn3.39}). We have found a closed-form classical solution to the primal HJB equation  for a highly complicated utility function $U$ in (\ref{example3.12}). 
}

\end{example}


\section{Primal and Dual Sufficient Conditions}

In this section we relax the differentiability condition of $V$.
Since $\psi(x)=V(e^x)$ is not assumed to be differentiable, we first rewrite (\ref{Po3}) in equivalent expressions which do not involve  derivatives of $\psi$.
\begin{eqnarray}
e^{\alpha x}w&=&\frac{1}{2\sqrt{\pi }}\int_{-\infty}^\infty
e^{-\frac{\eta^2}{4}-\alpha a\sqrt{t}\eta}\psi(x+a\sqrt{t}\eta)d\eta \label{Po3aa}\\
(e^{\alpha x}w)_x&=&\frac{1}{2\sqrt{\pi }}\int_{-\infty}^\infty
e^{-\frac{\eta^2}{4}-\alpha a\sqrt{t}\eta}\left({\eta\over 2a\sqrt{t}}+\alpha\right)\psi(x+a\sqrt{t}\eta)d\eta \label{Po3a}
\end{eqnarray}
Recall that the utility $U$ satisfies Assumption \ref{utilityturnpike}
and  the dual function $V$ is nonnegative, continuous, decreasing and convex on $R_+$.

\begin{theorem} \label{dualq}
Assume that  $V$ satisfies, for some $q<0$,
\begin{equation} \label{dualV1}
\lim_{y\rightarrow 0}\frac{V(y)}{y^q}=- \frac{1}{q}.
\end{equation}
Then  the turnpike property (\ref{turnpike}) holds.
If, in addition, there are positive constants $K$, $\alpha_1$ and $\delta$, such that
\begin{equation}\label{h-C2}
\left|\frac{V(y)}{y^q}+\frac{1}{q}\right|\leq K y^{\alpha_1},\ y\leq \delta,
\end{equation}
then the convergence rate (\ref{rate}) holds.
\end{theorem}

\noindent{\it Proof.}\
Since $V$ is not assumed to be differentiable, we cannot directly apply Theorem \ref{Main-1}. However, we know that for $t>0$, $v(t,y)$ is smooth decreasing strictly convex on $(0,\infty)$ and if the conditions (\ref{V-C}) and (\ref{V-C2}) for $v(t_0,y)$ hold for some $t_0>0$, then the turnpike property and convergent rate of the optimal investment hold. We next show that conditions (\ref{V-C}) and (\ref{V-C2}) hold for $v(t_0,y)$ at some $t_0>0$ according to
(\ref{dualV1}) and (\ref{h-C2}) respectively.

Suppose (\ref{dualV1}) holds. Let $\epsilon_0=-{1\over 2q}$. Then there exists a $\delta_0>0$ such that for $0<y<\delta_0$,
$0<V(y)<-{3\over 2q}y^q$. For $y\geq \delta_0$, $0\leq V(y)\leq V(\delta_0)$. Let $K_0=\max(-{3\over 2q}, V(\delta_0))$ and $x_0=\ln \delta_0$, we have
\begin{equation} \label{vex}
0\leq \psi(x):=V(e^x)\leq \left\{
\begin{array}{ll} K_0e^{qx}, & x<x_0,\\ K_0, & x\geq x_0.\end{array} \right.
\end{equation}

Using (\ref{Po3a}), we  get
\[
\left|\frac{(e^{\alpha x}w(t,x))_x}{e^{(q-\alpha)^2 a^2 t+qx}}+1\right|\leq \frac{1}{2 \sqrt{\pi}}\int_{-\infty}^\infty
e^{-\frac{(\eta-2(q-\alpha)a\sqrt{t})^2}{4}}\left|\frac{\eta}{2a\sqrt{t}}+\alpha\right|
\left|\frac{\psi(x+a\sqrt{t}\eta)}{e^{q(x+a\sqrt{t}\eta)}}+\frac{1}{q}\right|d\eta.
\]
By (\ref{vex}), we always have, if $x\leq 0$
\[
\left|\frac{\psi(x+a\sqrt{t}\eta)}{e^{q(x+a\sqrt{t}\eta)}}\right|\leq K_0(1+e^{-q a\sqrt{t}\eta }),
\]
for any $\eta$. Applying the dominated convergence theorem, we have
\[
\lim_{y\rightarrow 0}\left|\frac{v_y(t,y)}{e^{\lambda t} y^{q-1}} +1\right|
=\lim_{x\rightarrow -\infty}
\left|\frac{(e^{\alpha x}w(t,x))_x}{e^{(q-\alpha)^2 a^2 t+qx}}+1\right| =0.
\]
Hence
\[
\lim_{y\rightarrow 0}\left|\frac{v_y(t,y)}{ y^{q-1}} + e^{\lambda t}\right|
=0
\]
for any $t>0$. This proves (\ref{V-C}) for any  $t_0>0$ (see Remark \ref{remark3.6}).

To show the rate of convergence (\ref{rate}), we again only need to prove condition (\ref{V-C2}) holds.
Assume (\ref{h-C2}) holds. Set $x_0=\min(\ln \delta_0, \ln \delta)$ and
$K_0=\max(-{3\over 2q}, V(\delta_0), K)$.
Then $|\frac{\psi(x)}{e^{qx}}+{1\over q}|\leq K e^{\alpha_1 x}$
for $x\leq x_0$. For $x\leq 0$ we have
\begin{eqnarray*}
&&\left|\frac{(e^{\alpha x}w)_x}{e^{(q-\alpha)^2 a^2 t+qx}}+1\right|\\
&\leq&
  \frac{K_0}{2 \sqrt{\pi }}\int_{-\infty}^{\frac{x_0-x}{a\sqrt{t}}} \left|\frac{\eta}{2a\sqrt{t}}+\alpha\right|
e^{-\frac{(\eta+2(\alpha -q)a\sqrt{t})^2}{4}} e^{\alpha_1(x+a\sqrt{t}\eta)} d\eta\\
&& {}+ \frac{K_0}{2 \sqrt{\pi }}\int_{\frac{x_0-x}{a\sqrt{t}}}^\infty \left|\frac{\eta}{2a\sqrt{t}}+\alpha\right|
e^{-\frac{(\eta+2(\alpha -q)a\sqrt{t})^2}{4}}\left(1+\frac{1}{|q|}+e^{-q a\sqrt{t}\eta}\right)d\eta\\
&\leq& K_0(\alpha_1+|q|+\frac{1}{a\sqrt{\pi t}})e^{((\alpha -q-\alpha_1)^2-(\alpha -q)^2) a^2 t}e^{\alpha_1 x}
+\frac{K_0(1+|q|)}{2a|q|\sqrt{t}}M(\frac{x_0-x}{a\sqrt{t}}+2(\alpha -q)a\sqrt{t})\\
& &+\frac{K_0}{2a\sqrt{t}}e^{(\alpha^2 -(\alpha -q)^2) a^2 t}M(\frac{x_0-x}{a\sqrt{t}}+2\alpha a\sqrt{t})
+K_0(1+|q|)N(\frac{x_0-x}{a\sqrt{t}}+2(\alpha-q) a\sqrt{t})\\
&\leq& K_0 L_3(t)(e^{\alpha_1 x}+e^{\frac{x}{a\sqrt{t}}}),
\end{eqnarray*}
where
$L_3(t)=(\alpha_1+|q|+\frac{1}{a\sqrt{\pi t}})e^{((\alpha -q-\alpha_1)^2-(\alpha -q)^2) a^2 t}
+(1+|q|+\frac{1+2|q|}{2a|q|\sqrt{t}})e^{4- \frac{x_0}{a \sqrt{t}}}.$
We have used (\ref{N-M}) in the last inequality.  Note that
\[
\frac{v_y(t,y)}{ y^{q-1}} + e^{\lambda t}
=e^{\lambda t}[\frac{(e^{\alpha x}w(t,x))_x}{e^{(q-\alpha)^2 a^2 t+qx}}+1]
\]
We chose $t_0=\bar{t}$ and prove (\ref{V-C2}) (see Remark \ref{remark3.9}).
\qed


We give a condition on utility  $U$ that implies condition (\ref{dualV1}) and \ref{h-C2}.
\begin{corollary}\label{UtoV}
Assume that $U$ satisfies, for some $0<p<1$,
\begin{equation}\label{C3}
\lim_{x\rightarrow \infty}\frac{U(x)}{{1\over p}x^p}=1.
\end{equation}
Then the turnpike property
(\ref{turnpike}) holds.
If $U$ satisfies, for some $0<p<1, 0<\alpha<p, L>0, X>(\frac{L}{p}(p-\alpha))^{\frac{1}{\alpha}}$,
\begin{equation}\label{C31}
|\frac{U(x)}{{1\over p}x^p}-1|\leq Lx^{-\alpha},
\end{equation}
for $x\geq X$.
Then the convergence rate (\ref{rate}) holds  with $\alpha_1=\alpha/(1-p)$ and $q=p/(p-1)$.
\end{corollary}

\noindent{\it Proof.}\
We only need to show  (\ref{dualV1}), (\ref{h-C2}) hold with $q={p\over p-1}$. Theorem \ref{dualq} then gives the required conclusions.
Since $U$ is a concave function on $R_+$, the superdifferential of $U$ at $x\in R_+$ is a convex compact set, defined by
$$ \partial U(x)=\{\xi: U(y)\leq U(x)+\xi(y-x),\; y\in R_+\}.$$
The increasing property of $U$ and condition (\ref{C3}) imply that $\partial U(x)$ is a subset of $R_+$ and $0\not\in \partial U(x)$ for all $x\in R_+$. (If $0\in \partial U(\bar x)$ for some $\bar x\in R_+$, then  $U(x)\equiv U(\bar x)$ for all $x\geq \bar x$, which contradicts  (\ref{C3}).)
By (\ref{C3}), for any $0<\epsilon<1$,
there exists $X_\epsilon$, such that
\[
(1-\epsilon){1\over p}x^p\leq U(x) \leq (1+\epsilon){1\over p}x^p
\]
for $x\geq X_\epsilon$. The superdifferential of $U$ at $X_\epsilon$ is given by
$\partial U(X_\epsilon)=[a_\epsilon, b_\epsilon]$ for some $0<a_\epsilon\leq b_\epsilon<\infty$.
Note that $U(x)$ is concave. For $y<a_\epsilon$, we have, for $x\leq X_\epsilon$, that
$$ U(x)-xy\leq U(X_\epsilon)+a_\epsilon(x-X_\epsilon)-xy\leq U(X_\epsilon) - X_\epsilon y,$$
which implies that
\begin{eqnarray*}
V(y)&=&\max_{x\geq X_\epsilon}\{U(x)-xy\}\\
&\leq& \max_{x\geq X_\epsilon}\{(1+\epsilon){1\over p}x^p-xy\}\\
&= &-\frac{1}{q}(1+\epsilon)^{-\frac{1}{p-1}}y^q,
\end{eqnarray*}
provided $y<\min(a_\epsilon, (1+\epsilon)X_\epsilon^{p-1})$, where $q={p\over p-1}<0$. Dividing $y^q$ on both sides and letting $y\to0$ and then $\epsilon\to0$, we have
$$ \limsup_{y\to0} {V(y)\over y^q}\leq -{1\over q}.$$
Similarly, using $(1-\epsilon){1\over p}x^p\leq U(x)$ for $x\geq X_\epsilon$, we have
$$ \liminf_{y\to0} {V(y)\over y^q}\geq -{1\over q}.$$
We have proved (\ref{dualV1}).

Assume (\ref{C31}).
Let $\partial U(X)=[a, b]$ for some $0<a \leq b<\infty$. For $y<a$, $x\leq X$, we have
$U(x)-xy\leq U(X)- Xy$.
Hence we obtain, for $y<\min\{a, X^{p-1}+\frac{L}{p}(p-\alpha)X^{p-1-\alpha}\}$
\begin{eqnarray*}
V(y)&=&\max_{x\geq X}\{U(x)-xy\}\\
&\leq& \max_{x\geq X}\{(1+Lx^{-\alpha}){1\over p}x^p-xy\}\\
&= &-\frac{1}{q}x(y)^p+\frac{L}{p}(1+\alpha-p)x(y)^{p-\alpha},
\end{eqnarray*}
where
\[
y=x(y)^{p-1}+\frac{L}{p}(p-\alpha)x(y)^{p-1-\alpha},
\]
and $x(y)\geq X$.
From this relation, we conclude that $x(y)^{-\alpha}\leq y^{\frac{\alpha}{1-p}}$ and
\begin{eqnarray*}
\frac{x(y)^p}{y^q}&=&(1+\frac{L}{p}(p-\alpha)x(y)^{-\alpha})^{-q}\\
&\leq& 1+\frac{L}{1-p}(p-\alpha)(1+\frac{L}{p}(p-\alpha)X^{-\alpha})^{-q}x(y)^{-\alpha}.
\end{eqnarray*}
Hence,
\begin{eqnarray*}
\frac{V(y)}{y^q}+\frac{1}{q} &\leq& -\frac{1}{q}(\frac{x(y)^p}{y^q}-1)+\frac{L}{p}(1+\alpha-p)\frac{x(y)^{p-\alpha}}{y^q}\\
&\leq& \biggl(\frac{L}{p}(p-\alpha)(1+\frac{L}{p}(p-\alpha)X^{-\alpha})^{-q}\\
&& {} +\frac{L}{p}(1+\alpha-p)[1+\frac{L(p-\alpha)}{1-p}(1+\frac{L}{p}(p-\alpha)X^{-\alpha})^{-q}]\biggr)x(y)^{-\alpha}\\
&\leq& \frac{L}{p}(1+\frac{L}{p}(p-\alpha)X^{-\alpha})^{-q} [(p-\alpha)
+(1+\alpha-p)(1+\frac{L(p-\alpha)}{1-p})]y^{\frac{\alpha}{1-p}}.
\end{eqnarray*}
The lower bound of $\frac{V(y)}{y^q}+\frac{1}{q}$ can be derived similarly.
Let $\alpha_1=\alpha/(1-p)$. We have  (\ref{h-C2}).
\qed

\begin{remark}{\rm
 (\ref{C3}) can be replaced by $\lim_{x\to\infty} U(x)/x^p=k$ for some positive constant $k$ due to the invariance property of the optimal trading strategy for a scaled objective function (see Remark \ref{remark3.6} and Corollary \ref{corollary3.7}). In this case (\ref{C31}) is replaced by $|U(x)/x^p-k|\leq Lx^{-\alpha}$ for $x\geq X$.
There exists a wider class of utility functions $U$ that satisfy (\ref{C3}) and (\ref{C31}), for example, $U$ defined in  (\ref{example3.12}) satisfies both  conditions with $p=3/4$, $k=1$ and $\alpha=1/2$, which can be easily seen from the observation that as $x\to\infty$, $H(x)\sim x^{-1/4}$ and $U(x)\sim (4/3)x^{3/4} + x^{1/4}$.  In fact, if we define
$U(x)=U_1(x)$ for $0\leq x<K$ and $x^p/p$ for $x\geq K$, where $0<p<1$, $K>0$ and $U_1$ is a function such that $U$ satisfies
Assumption \ref{utility}.  There are infinitely many such functions $U_1$.  The reason for this   is that (\ref{C3}) and (\ref{C31}) only place the restriction on the limiting behavior of the utility when wealth level is very high but leave the freedom  for  other levels of wealth.
}
\end{remark}

\begin{theorem}\label{dual0}
Assume that  $V$ satisfies
\begin{equation}\label{duallog}
\lim_{y\rightarrow 0}\frac{V(y)}{\ln y}=-1.
\end{equation}
Then  the turnpike property (\ref{turnpike}) holds.
If, in addition, there are positive constants $K$, $\alpha_1$ and $\delta$ such that
\begin{equation}\label{h-C2log}
|V(y)+ \ln y|\leq K y^{\alpha_1},\ y\leq \delta,
\end{equation}
then the convergence rate (\ref{rate}) holds.
\end{theorem}

\noindent{\it Proof.}\ We need to show the conditions (\ref{V-C}) and (\ref{V-C2}) for $v(t_0,y)$ hold for some $t_0>0$ with $q=0$.
Assume (\ref{duallog}).
Let $\epsilon_0={1\over 2}$. Then there exists a $\bar\delta>0$, such that for $0<y<\bar\delta$, $0<-{1\over 2}\ln y< V(y)< -{3\over 2}\ln y$. Let $\delta_0=\min(\bar\delta, 1)$. For $y\geq \delta_0$,
$0\leq V(y)\leq V(\delta_0)$. Let $K_0=\max({3\over 2}, V(\delta_0))$ and $x_0=\ln \delta_0\leq 0$, we have
\begin{equation} \label{vex0}
0\leq \psi(x):=V(e^x)\leq \left\{
\begin{array}{ll} -K_0x, & x<x_0,\\ K_0, & x\geq x_0.\end{array} \right.
\end{equation}
We have, by (\ref{Po3aa})
\[
\frac{e^{\alpha x + \beta t}w(t,x)}{x}+1 =\frac{1}{2\sqrt{\pi }}\int_{-\infty}^\infty
e^{-\frac{1}{4}(\eta +2\alpha a\sqrt{t})^2}\left(\frac{\psi(x+a\sqrt{t}\eta)}{x}+1\right)d\eta.
\]
For $x\leq -1$, we always have
\[
\left|\frac{\psi(x+a\sqrt{t}\eta)}{x}\right|\leq K_0(1+a\sqrt{t}|\eta|), \ \forall \eta.
\]
The dominated convergence theorem gives
\[
\lim_{y\rightarrow 0}\frac{v(t,y)}{\ln y}=\lim_{x\rightarrow -\infty}\frac{e^{\alpha x+ \beta t}w(t,x)}{x}=-1.
\]
This implies
\[
\lim_{y\rightarrow 0}\frac{v_y(t,y)}{y^{-1}}=-1
\]
for any $t>0$.

Assume, in addition, (\ref{h-C2log}) holds. Set $\delta_0=\min(\bar\delta, 1, \delta)$ and
$K_0=\max({3\over 2}, V(\delta_0), K)$, then, for $x\leq x_0:=\ln \delta_0$,
$|\psi(x)+x|\leq K_0 e^{\alpha_1 x}$, also note that
$$
e^{\beta t}(e^{\alpha x}w)_x+1=
 \frac{1}{2 \sqrt{\pi }}\int_{-\infty}^\infty
\left(\frac{\eta}{2a\sqrt{t}}+\alpha\right)
e^{-\frac{(\eta+2\alpha a\sqrt{t})^2}{4}}
\left(\psi(x+a\sqrt{t}\eta)+(x+a\sqrt{t}\eta)\right)d\eta.
$$
If $\eta\geq\frac{x_0-x}{a\sqrt{t}}$, then
We have
\[
|x+a\sqrt{t}\eta|\leq|x_0|+x-x_0+a\sqrt{t}\eta\leq 2|x_0|+a\sqrt{t}\eta
\]
for $x\leq 0$. Hence
\begin{eqnarray*}
|e^{\beta t}(e^{\alpha x}w)_x+1|
&\leq&
 \frac{K_0}{2 \sqrt{\pi }}\bigg(\int_{-\infty}^{\frac{x_0-x}{a\sqrt{t}}} \left|\frac{\eta}{2a\sqrt{t}}+\alpha\right|
e^{-\frac{(\eta+2\alpha a\sqrt{t})^2}{4}} e^{\alpha_1(x+a\sqrt{t}\eta)} d\eta\\
&&{}+ \int_{\frac{x_0-x}{a\sqrt{t}}}^\infty \left|\frac{\eta}{2a\sqrt{t}}+\alpha\right|
e^{-\frac{(\eta+2\alpha a\sqrt{t})^2}{4}}(1+2|x_0|+ a\sqrt{t}\eta)d\eta\bigg) \\
&\leq& K L_4(t)\left(e^{\alpha_1 x}+e^{\frac{x}{a\sqrt{t}}}\right)
\end{eqnarray*}
for some $L_4(t)$. We chose $t_0=\bar{t}$ and complete the proof.
\qed

We give a condition on utility  $U$ that implies condition (\ref{duallog}).
\begin{corollary}\label{UtoV0}
Assume that  $U$ satisfies
\begin{equation}\label{C3log}
\lim_{x\rightarrow \infty}\frac{U(x)}{\ln x}=1.
\end{equation}
Then the turnpike property (\ref{turnpike}) holds with $p=0$.§
If $U$ satisfies, for some $\alpha>0, L>0, X>(\alpha L)^{\frac{1}{\alpha}}$,
\begin{equation}\label{C3-1log}
|U(x)-\ln x -1|\leq Lx^{-\alpha}
\end{equation}
for $x\geq X$.
Then the convergence rate (\ref{rate}) holds.
\end{corollary}

\noindent{\it Proof.}\
The proof  is the same as that of Corollary \ref{UtoV}. The difference is to replace ${1\over p}x^p$ by $\ln x$.
\qed

\begin{theorem}
Assume that $V$ satisfies, for some $0<q<1$,
\begin{equation}\label{dualfinite}
V(0)<\infty, \ \
\lim_{y\rightarrow 0}\frac{V(y)-V(0)}{y^q}= -\frac{1}{q}.
\end{equation}
Then  the turnpike property (\ref{turnpike}) holds.
If, in addition, there are positive constants $K$, $\alpha_1$ and $\delta$, such that
\begin{equation}\label{h-C3}
\left|\frac{V(y)-V(0)}{y^q}+\frac{1}{q}\right|\leq K y^{\alpha_1},\ y\leq \delta,
\end{equation}
then the convergence rate  (\ref{rate}) holds.
\end{theorem}

\noindent{\it Proof.}\ We only need to show (\ref{V-C}), (\ref{V-C2}) for $v(t_0,y)$ hold for some $t_0>0$ and $0<q<1$.
Assume (\ref{dualfinite}).
Since $V(0)<\infty$ and $V$ is a nonnegative decreasing function, we have $0\leq \psi(x)=V(e^x)\leq V(0)$ for all $x$. A simple calculus shows that
\[
\frac{1}{2\sqrt{\pi }}\int_{-\infty}^\infty
e^{-\frac{(\eta-2(q-\alpha)a\sqrt{t})^2}{4}}\left(\frac{\eta}{2a\sqrt{t}}+\alpha\right)
{V(0)\over e^{q(x+a\sqrt{t}\eta)}} d\eta=0
\]
and
\[
\frac{1}{2\sqrt{\pi }}\int_{-\infty}^\infty
e^{-\frac{(\eta-2(q-\alpha)a\sqrt{t})^2}{4}}\left(\frac{\eta}{2a\sqrt{t}}+\alpha\right)
{1\over q} d\eta=1.
\]
Hence
\begin{eqnarray*}
\frac{(e^{\alpha x}w(t,x))_x}{e^{(q-\alpha)^2 a^2 t+qx}}+1
&=& \frac{1}{2 \sqrt{\pi}}\int_{-\infty}^\infty
e^{-\frac{(\eta-2(q-\alpha)a\sqrt{t})^2}{4}}\left(\frac{\eta}{2a\sqrt{t}}+\alpha\right)
\left(\frac{\psi(x+a\sqrt{t}\eta)-V(0)}{e^{q(x+a\sqrt{t}\eta)}}+\frac{1}{q}\right)d\eta.
\end{eqnarray*}
Condition (\ref{dualfinite}) implies that for a fixed $\epsilon_0>0$ there exists $X_0$ such that for all $x< X_0$,
$$ \left|{\psi(x)-V(0)\over e^{qx}} + {1\over q}\right| <\epsilon_0.$$
Therefore, for any $\eta$, if $x+a\sqrt{t}\eta<X_0$, then
$$  \left|{\psi(x+a\sqrt{t}\eta)-V(0)\over e^{q(x+a\sqrt{t}\eta)}} + {1\over q}\right| <\epsilon_0$$
and if $x+a\sqrt{t}\eta\geq X_0$, then
$$  \left|{\psi(x+a\sqrt{t}\eta)-V(0)\over e^{q(x+a\sqrt{t}\eta)}} + {1\over q}\right|
\leq 2V(0)e^{-q(x+a\sqrt{t}\eta)}+{1\over q}\leq 2V(0)e^{-qX_0}+{1\over q}.$$
The estimates above, the dominated convergence theorem and  (\ref{dualfinite})
give the required limit:
$$ \lim_{y\to0} {v_y(t,y)\over e^{\lambda t} y^{q-1}} =
\lim_{x\to-\infty}\frac{(e^{\alpha x}w(t,x))_x}{e^{(q-\alpha)^2 a^2 t+qx}}=-1.$$
We have proved the turnpike property (\ref{V-C}).

If, in addition, (\ref{h-C3}) holds,
then setting $x_0=\min(\ln \delta, X_0)$ and $K_0=\max(K, \epsilon_0,
2V(0)e^{-qX_0}+{1\over q})$,  we have
\begin{eqnarray*}
\left|\frac{(e^{\alpha x}w)_x}{e^{(q-\alpha)^2 a^2 t+qx}}+1\right|
&\leq& \frac{1}{2 \sqrt{\pi}}\int_{-\infty}^{{x_0-x\over a\sqrt{t}}}
e^{-\frac{(\eta-2(q-\alpha)a\sqrt{t})^2}{4}}\left|\frac{\eta}{2a\sqrt{t}}+\alpha\right|
K_0e^{\alpha_1(x+a\sqrt{t}\eta)} d\eta\\
&& {}+ \frac{1}{2 \sqrt{\pi}}\int^{\infty}_{{x_0-x\over a\sqrt{t}}}
e^{-\frac{(\eta-2(q-\alpha)a\sqrt{t})^2}{4}}\left|\frac{\eta}{2a\sqrt{t}}+\alpha\right|K_0d\eta\\
&\leq& L_5(t)(e^{\alpha_1 x}+e^{\frac{x}{a\sqrt{t}}}),
\end{eqnarray*}
for some $L_5(t)$, which implies (\ref{V-C2}).  We have shown the rate of convergence (\ref{rate}).
\qed

We give a condition on utility  $U$ that implies condition (\ref{dualfinite}).
\begin{corollary}\label{UtoVfinite}
Assume that $U$ satisfies, for some $p<0$,
\begin{equation}\label{C3finite}
U(\infty)<\infty, \ \ \lim_{x\rightarrow \infty}\frac{U(x)-U(\infty)}{{1\over p}x^p}=1.
\end{equation}
Then the turnpike property (\ref{turnpike}) holds.
If $U$ satisfies, for some $p<0, \alpha>0, L>0, X>(\frac{L(p-\alpha)(p-\alpha-1)}{p(p-1)})^{\frac{1}{\alpha}}$,
$$
|\frac{U(x)-U(\infty)}{{1\over p}x^p}-1|\leq Lx^{-\alpha},
$$
for $x\geq X$.
Then the convergence rate (\ref{rate}) holds.
\end{corollary}
\noindent{\it Proof.}\
The proof  is the same as that of Corollary \ref{UtoV}. We only need to note that
$V(0)=U(\infty)$ and
$(1+\epsilon){1\over p}x^p\leq U(x)-U(\infty) \leq (1-\epsilon){1\over p}x^p$ for $x\geq X_\epsilon$ due to $p<0$. The rest follows the same lines of reasoning.
\qed

\begin{remark}{\rm  \label{remark4.8} We have given a number of sufficient conditions that guarantee the turnpike property and the convergence rate. If $U$ does not satisfy these conditions then we cannot directly conclude if the turnpike property holds or not and need to use other methods to check it.
One example is $U(x)=x\wedge H$ which is not strictly increasing when $x\geq H$, a condition required for the turnpike property. From Example \ref{ex1}, we know the optimal amount of investment in the risky asset is equal to (see (\ref{eqn2.26})  and recall that $t$ is the time to horizon in this section)
$$
A(t,x)={He^{-rt}\over \sigma\sqrt{t} }\phi\left(\Phi^{-1}({x\over H}e^{rt})\right)
$$
if $xe^{rt}\leq H$ and 0 otherwise. Therefore  $A(t,x)=0$}
 as $t\to\infty$ and the turnpike property does not hold.  Another example is $U(x)=1-e^{- x}$  that does not satisfy (\ref{C3finite}).  We show in the next example that  the turnpike property does not hold.
\end{remark}

\begin{example}{\rm  \label{ex5}
Assume $U(x)=1-e^{-x}$ for $x\geq0$ and $-\infty$ for $x<0$. The dual function of $U$ is given by $V(y)=(1+y(\ln y-1))1_{\{0<y\leq1\}}$.
We have $V''(y)=y^{-1}1_{\{0<y<1\}}$. From (\ref{kac}) we find that
$$v_{yy}(t,y)=E[V''(y\tilde Y)\tilde Y^2]=E[{1\over y}1_{\{y\tilde Y<1\}} \tilde Y]$$
where $\tilde Y=\exp\left(-(r+{1\over 2}\theta^2)t -\theta \sqrt{t}Z\right)$ and
$Z$ is a standard normal variable.
This leads to
$$A(t,x)={\theta\over\sigma} yv_{yy}(t,y)
={\theta\over\sigma} E[1_{\{y\tilde Y<1\}} \tilde Y].$$
Finally, due to the equivalence of  $y\tilde Y<1$ and $Z>k:={1\over \theta\sqrt{t}}(\ln y-(r+{1\over 2}\theta^2)t)$, we have
$$ A(t,x)={\theta\over\sigma}\int_{k}^\infty
e^{-(r+{1\over 2}\theta^2)t -\theta \sqrt{t}z} {1\over \sqrt{2\pi}}e^{-{z^2\over 2}}dz
={\theta\over\sigma}e^{-rt}\Phi(-k-\theta\sqrt{t})$$
which tends to 0 as $t\to\infty$. The turnpike property does not hold.
Note that  $U$ is not an exponential utility function in the usual sense as it is only defined on the positive real line, not on the whole real line. The optimal portfolio $A(t,x)$ depends on both $t$ and  $x$ and is not a function of  $t$ only as in the case of a standard exponential utility function. Note also that the relative risk aversion coefficient of $U$ is $R(x)=x$, an increasing function, which shows that $U$ is a HARA utility  representing an investor who will decrease the percentage of wealth invested in the risky asset as wealth increases, such economic behavior clearly violates the turnpike property.

}

\end{example}

\section{Conclusions}
In this paper we  discuss the turnpike property and the convergence rate of a long term investor with a power-like utility for large wealth. We first extend the results of Bian et al. (2011) to more general utilities  and show constructively the existence of a smooth solution to the HJB equation for the investment problem. We demonstrate the usefulness of the result by solving a terminal wealth maximization problem and providing a  closed-form smooth solution to the HJB equation.
We then prove the main results of the paper on the turnpike property and the convergence rate when the dual function of the utility is  differentiable and its derivative satisfies some growth and limiting conditions. We illustrate these results with a nontrivial example. We finally list some sufficient conditions that guarantee  the turnpike property and the convergence rate  in terms of both the utility function and its dual function  while removing the usual assumptions of the differentiability and the strict concavity of the utility function. 
As commented by the reviewer, the assessment of the turnpike property via the dual value function depends exclusively on the structure of the wealth process and the geometric Brownian motion asset price process in this paper. It would be interesting to see if the turnpike property and the convergence rate still hold for more general asset price processes such as the stochastic volatility and L\'evy processes. These open questions require further research and investigation.

\bigskip\noindent
{\bf Acknowledgment}. The authors are very grateful to  the anonymous reviewer whose constructive comments and suggestions have helped to improve the paper of the previous two versions.


\end{document}